\documentclass{article}
\usepackage{amssymb,amsmath,amsthm,times}
\usepackage{color,graphicx}
\usepackage{epstopdf}
\usepackage[numbers,sort&compress]{natbib}
\newcommand{\comment}[1]{}   

\def\d{{\rm d}}

\def\D{\mbox{\rm D}}  
\def\i{\mbox{\rm i}}
\def\cdot{{\scriptstyle\,\bullet\,}}

\def\p{Pain\-lev\'{e}}
\def\kp{Kadomtsev-Petviashvili}
\def\sch{Schr\"{o}dinger}
\def\PI{\mbox{\rm P$_{\rm I}$}}
\def\PII{\mbox{\rm P$_{\rm II}$}}

\def\PIV{\mbox{\rm P$_{\rm IV}$}}

\newcommand{\pderiv}[3][]{\frac{\partial^{#1}{#2}}{{\partial{#3}}^{#1}}}

\def\a{\alpha}
\def\b{\beta}

\def\la{\lambda}
\def\etal{\textit{et al.}}
\def\i{\ifmmode{\rm i}\else\char"10\fi}

\newtheorem{theorem}{Theorem}

\newtheorem{remark}[theorem]{Remark}

\newtheorem{conjecture}[theorem]{Conjecture}
\newtheorem{lemma}[theorem]{Lemma}
\numberwithin{equation}{section}
\numberwithin{table}{section}
\numberwithin{figure}{section}
\numberwithin{theorem}{section}

\begin{document}
\title{Rational solutions of the Boussinesq equation\\ and applications to rogue waves}
\author{Peter A.\ Clarkson and Ellen Dowie\\ School of Mathematics, Statistics and Actuarial Science\\
University of Kent, Canterbury, CT2 7NF, UK\\
Email: \texttt{P.A.Clarkson@kent.ac.uk}, \texttt{ed275@kent.ac.uk}
}

\maketitle
\begin{abstract}
We study rational solutions of the Boussinesq equation, which is a soliton equation solvable by the inverse scattering method. These rational solutions, which are algebraically decaying and depend on two arbitrary parameters, are expressed in terms of special polynomials that are derived through a bilinear equation, have a similar appearance to rogue-wave solutions of the focusing nonlinear \sch\ (NLS) equation. Further the rational solutions have an interesting structure as they are comprised of a linear combination of four independent solutions of the bilinear equation. 
Rational solutions of the \kp\ I (KPI) equation are derived in two ways, from rational solutions of the NLS equation and from rational solutions of the Boussinesq equation. It is shown that these two families of rational solutions of the KPI equation are fundamentally different and a unifying framework is found which incorporates both families of solutions.
\end{abstract}

\section{\label{sec1}Introduction} 
``Rogue waves", sometimes knows as ``freak waves" or ``monster waves", are waves appearing as extremely large, localized waves in the ocean which have been of considerable interest recently, cf.~\citep{refDKM,refKPS,refOsborne,refPelKha}. The average height of rogue waves is at least twice the height of the surrounding waves, are very unpredictable and so they can be quite unexpected and mysterious. A feature of rogue waves is that they ``come from nowhere and disappear with no trace" \citep{refAAT,refASGA}. 
In recent years, the concept of rogue waves has been extended beyond oceanic waves: to pulses emerging from optical fibres \citep{refDudley,refDudleyDEG,refKibler,refSolli}; waves in Bose-Einstein condensates \citep{refBKA}; in superfluids \citep{refGEKLKK}; in optical cavities \citep{refMBRA}, in the atmosphere \citep{refStenMark}; and in finance \citep{refYan10,refYan11}; for a comprehensive review of the different physical contexts rogue waves arise see \citep{refORBMA}.
The most commonly used mathematical model for rogue waves involves rational solutions of the focusing nonlinear \sch\ (NLS) equation    \begin{equation}
\mbox{\rm i} \psi_{t} + \psi_{xx}+\tfrac1{2}|\psi|^2 \psi =0,
\label{eq:fnls}
\end{equation}
where subscripts denote partial derivatives, with $\psi$ the wave envelope, $t$ the temporal variable and $x$ the spatial variable in the frame moving with the wave, see \S\ref{sec2}. 

In this paper we are concerned with rational solutions of the Boussinesq equation 
\begin{equation}
u_{tt}+u_{xx}-(u^2)_{xx}-\tfrac13u_{xxxx}=0,\label{eq:bq}
\end{equation}
which are algebraically decaying and have a similar appearance to rogue-wave solutions of the NLS equation \eqref{eq:fnls}.
Equation \eqref{eq:bq} was introduced by Boussinesq in 1871 to describe the propagation of long waves in shallow water \citep{refBousa,refBousc}; see, also \citep{refUrs,refWhit}. 
The Boussinesq equation \eqref{eq:bq} is also a soliton equation solvable by inverse scattering \citep{refAC,refAH,refAS81,refDTT,refZak} which arises in several other physical applications including one-dimensional nonlinear lattice-waves \citep{refToda,refZab}; vibrations in a nonlinear string \citep{refZak}; and ion sound waves in a plasma \citep{refInR,refScott}. We remark that equation \eqref{eq:bq} is sometimes referred to as the ``bad" Boussinesq equation, i.e.\ when the ratio of the $u_{tt}$ and $u_{xxxx}$ terms is negative. If the sign of the $u_{xxxx}$ term is reversed in \eqref{eq:bq}, then the equation is sometimes called the ``good" Boussinesq equation. The coefficients of the $u_{xx}$ and $(u^2)_{xx}$ terms can be changed by scaling and translation of the dependent variable $u$. For example, letting $u\to u+1$ in \eqref{eq:bq} gives
\begin{equation}
u_{tt}-u_{xx}-(u^2)_{xx}-\tfrac13u_{xxxx}=0,\label{eq:bq1}
\end{equation}
which is the non-dimensionalised form of the equation originally written down by Boussinesq  \cite{refBousa,refBousc}.

There has been considerable interest in partial differential equations solvable by inverse scattering, the \textit{soliton equations}, since the discovery in 1967 by Gardner, Greene, Kruskal and Miura \citep{refGGKM} of the method for solving the initial value problem for the Korteweg-de Vries (KdV) equation
\begin{equation} u_t + 6uu_x + u_{xxx} =0.\label{eq:kdv} \end{equation} 
During the past forty years or so there has been much interest in rational solutions of the soliton equations. For some soliton
equations, solitons are given by rational solutions, e.g.\ for the Benjamin-Ono equation \citep{refMatsuno79,refSatIsh79}
Further applications of rational solutions to soliton equations include:
in the description of vortex dynamics \citep{refAref07a,refAref07b,refANSTV};
vortex solutions of the complex sine-Gordon equation \citep{refBP98,refOB05};
and in the transition behaviour for the semi-classical sine-Gordon equation \citep{refBM12}.

In \S\ref{sec2}, we discuss rational solutions of the focusing NLS equation \eqref{eq:fnls}, including some generalised rational solutions which involve two arbitrary parameters. 
In \S\ref{sec3}, we discuss rational solutions of the Boussinesq equation \eqref{eq:bq}, also including some generalised rational solutions which involve two arbitrary parameters. Further the generalised rational solutions have an interesting structure as they are comprised of a linear combination of four independent solutions of an associated bilinear equation. 
In \S\ref{sec4}, we discuss rational solutions of the \kp\ I (KPI) equation 
\begin{equation}\label{eq:kp1}
(v_\tau+6vv_\xi+v_{\xi\xi\xi})_\xi=3 v_{\eta\eta},
\end{equation}
which are derived in two ways, first from rational solutions of the focusing NLS equation \eqref{eq:fnls} and second from rational solutions of the Boussinesq equation \eqref{eq:bq}. In the simplest nontrivial case, it is shown that these two types of rational solutions are different. Further we derive a more general rational solution which has those related to the focusing NLS and Boussinesq equations as special cases and so provides a unifying framework. In \S\ref{sec5} we discuss our results.

\section{\label{sec2}Rational solutions of the focusing nonlinear \sch\ equation}
The nonlinear \sch\ (NLS) equation
\begin{equation}
\mbox{\rm i} \psi_{t} + \psi_{xx}+\tfrac1{2}\sigma |\psi|^2 \psi =0,\qquad\sigma=\pm1,
\label{eq:nls}
\end{equation}
is one of the most important nonlinear partial differential equations. In 1972, Zakharov and Shabat \citep{refZS72} developed the inverse scattering method of solution for it.  Prior to the discovery that the NLS equation \eqref{eq:nls} was solvable by inverse scattering, it had been considered by researchers in water waves \citep{refBN,refBR,refZakh68} (see also \citep{refMJAbook,refAS79,refAS81}). In 1973, Hasegawa and Tappert \citep{refHTa,refHTb} discussed the relevance of the NLS equation \eqref{eq:nls} in optical fibres and their associated solitary wave solutions. Hasegawa and Tappert showed that optical fibres could sustain envelope solitons -- both bright and dark solitons. Bright solitons, which decay as $|x|\to\infty$, arise with anomalous (positive) dispersion for \eqref{eq:nls} with $\sigma=1$, the \emph{focusing NLS equation}. Dark solitons, which do not decay as $|x|\to\infty$, arise with normal (negative) dispersion for \eqref{eq:nls} with $\sigma=-1$, the \emph{de-focusing NLS equation}. 

Rational solutions of the focusing NLS equation \eqref{eq:fnls} have the general form
\begin{equation}
\psi_n(x,t)=\left\{1-4\frac{G_n(x,t)+{\rm i} tH_n(x,t)}{{D}_n(x,t)}\right\}\exp\left(\tfrac12{\rm i} t\right),
\end{equation}
where $G_n(x,t)$ and $H_n(x,t)$ are polynomials of degree $\tfrac12(n+2)(n-1)$ in both $x^2$ and $t^2$, with total degree $\tfrac12(n+2)(n-1)$, and ${D}_n(x,t)$ is a polynomial of degree $\tfrac12n(n+1)$ in both $x^2$ and $t^2$, with total degree $\tfrac12n(n+1)$ and has no real zeros. The polynomials $D_n(x,t)$, $G_n(x,t)$ and  $H_n(x,t)$ satisfy the Hirota equations
\begin{align*}
&4(t\D_t+1) H_n\cdot D_n+\D_x^2 D_n\cdot D_n - 4\D_x^2D_n\cdot G_n=0,\\
&\D_t G_n\cdot D_n+t\D_x^2H_n\cdot D_n=0,\\
&\D_x^2D_n\cdot D_n=8G_n^2+8t^2H_n^2-4 D_nG_n,
\end{align*}
with $\D_x$ and $\D_t$ the Hirota operators
\begin{align}\label{hirotaop}
\D_x^\ell \,\D_t^m F(x,t)\cdot F(x,t)&=\left[\left(\pderiv{}{x}-\pderiv{}{x'}\right)^\ell\left(\pderiv{}{t}-\pderiv{}{t'}\right)^m F(x,t)F(x',t')\right]_{x'=x,t'=t}.
\end{align}
The first two rational solutions of the focusing NLS equation \eqref{eq:fnls} have the form \citep{refAASG,refACA}
\begin{align}
\psi_1(x,t)&=\left\{1-\frac{4(1+{\rm i} t)}{x^2+t^2+1}\right\}\exp\left(\tfrac12{\rm i} t\right),\label{nls:psi1}\\
\psi_2(x,t)&=\left\{1-12\,\frac{G_2(x,t)+{\rm i} tH_2(x,t)}{{D}_2(x,t)}\right\}\exp\left(\tfrac12{\rm i} t\right),\label{nls:psi2}
\end{align}
where 
\begin{subequations}
\begin{align}
G_2(x,t)&= x^4 + 6(t^2+1)x^2+ 5t^4+18t^2-3,\\
H_2(x,t)&= x^4+2(t^2-3)x^2+ (t^2+5)(t^2-3),\\
{D}_2(x,t)&= x^6+3(t^2+1)x^4 + 3(t^2-3)^2x^2 + t^6+27t^4+99t^2+9, \label{nls:f2}
\end{align}\end{subequations}
The solution $\psi_1(x,t)$ given by \eqref{nls:psi1} is known as the ``Peregrine solution" \citep{refPeregrine}. Further
\[|\psi_n(x,t)|^2=1+4\pderiv[2]{}{x}\ln D_n(x,t).\]

\begin{figure}[ht]
\[\begin{array}{c@{\quad}c@{\quad}c} 
\includegraphics[width=2in]{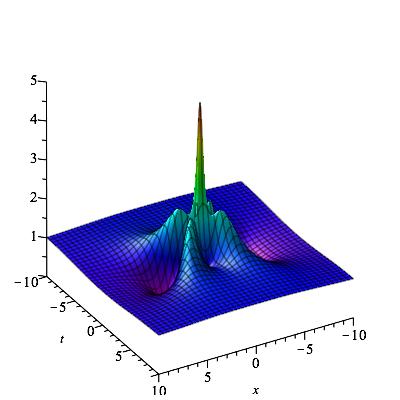}  & \includegraphics[width=2in]{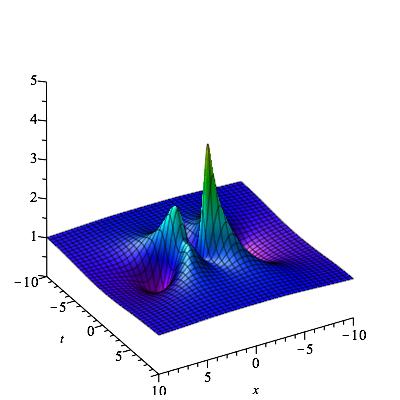} &
\includegraphics[width=2in]{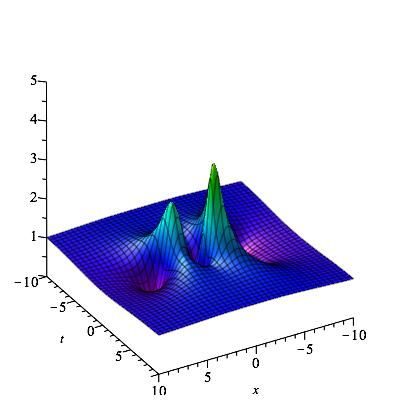} \\
\alpha=\beta=0 &  \alpha=\beta=5 &  \alpha=\beta=10\\
\includegraphics[width=2in]{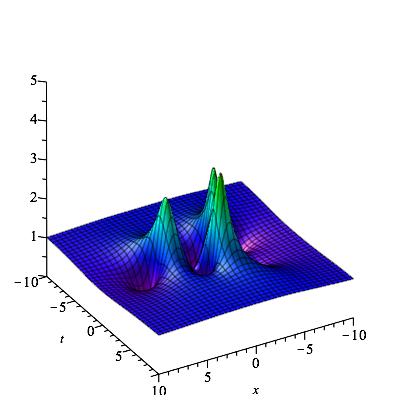}& 
 \includegraphics[width=2in]{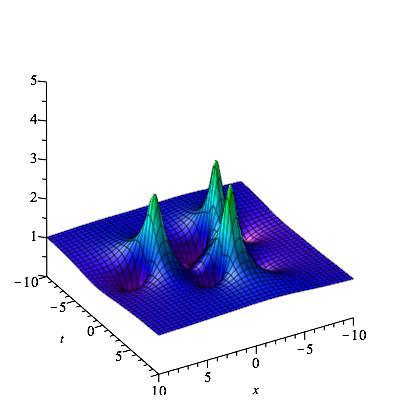}& \includegraphics[width=2in]{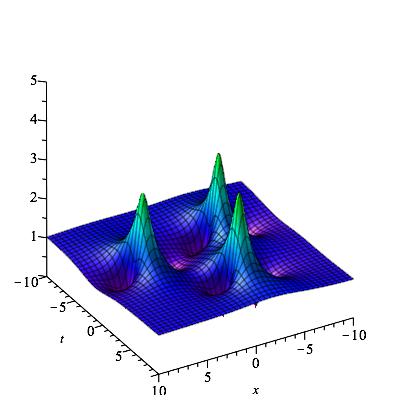}\\
  \alpha=\beta=20  & \alpha=\beta=50  & \alpha=\beta=100 
 \end{array}\]
\caption{\label{fig:nlsu2gen}Plots of the generalised rational solution $|\widehat{\psi}_2(x,t;\alpha,\beta)|$ given by \eqref{nls:u2hat} of the focusing NLS equation for various values of the parameters $\alpha$ and $\beta$.}
\end{figure}

Dubard \etal\ \citep{refDGKM} show that the rational solutions of the focusing NLS equation \eqref{eq:fnls} can be generalised to include some arbitrary parameters. The first of these generalized solutions has the form
\begin{align}\label{nls:u2hat}\widehat{\psi}_2(x,t;\alpha,\beta)=
\left\{1-12\frac{\widehat{G}_2(x,t;\alpha,\beta)+{\rm i} \widehat{H}_2(x,t;\alpha,\beta)}{\widehat{D}_2(x,t;\alpha,\beta)}\right\}\exp\left(\tfrac12{\rm i} t\right),\end{align}
where
\begin{subequations}\begin{align}
\widehat{G}_2(x,t;\alpha,\beta)&= 
G_2(x,t) - 2\a t + 2\beta x,\\
\widehat{H}_2(x,t;\alpha,\beta)&= 
tH_2(x,t)+ \a(x^2-t^2+1) + 2\beta xt,\\
\widehat{D}_2(x,t;\alpha,\beta)&= 
{D}_2(x,t) + 2\a t(3x^2-t^2-9) - 2\beta x(x^2-3t^2-3)+\a^2+\beta^2,\label{nls:F2}
\end{align}\end{subequations}
with $\alpha$ and $\beta$ arbitrary constants, see also \citep{refDubMat11,refDubMat13,refKAA11,refKAA12,refKAA13}. 
These generalized solutions have now been expressed in terms of Wronskians, see Gaillard \citep{refGail11,refGail12,refGail13a,refGail13b,refGail13c,refGail13d,refGail13e,refGail13f,%
refGail14a,refGail14b,refGail14c,refGail14d,refGail15a,refGail15b,refGail15c,refGailGas1,refGailGas2,refGailGas3}, Guo, Ling and Liu \citep{refGLL}, Ohta and Yang \citep{refOY}. We note that the polynomial $\widehat{D}_2(x,t;\alpha,\beta)$ has the form
\begin{equation}\label{nls:F2hat} \widehat{D}_2(x,t;\alpha,\beta)={D}_2(x,t) +2\a tP_1(x,t)+2\beta xQ_1(x,t)+\a^2+\beta^2,\end{equation}
where $P_1(x,t)$ and $Q_1(x,t)$ are linear functions of $x^2$ and $t^2$.
In Figure \ref{fig:nlsu2gen}, plots of the generalised rational solution $|\widehat{\psi}_2(x,t;\alpha,\beta)|$ given by \eqref{nls:u2hat} of the focusing NLS equation for various values of the parameters $\alpha$ and $\beta$. The solution has a single peak when $\alpha=\beta=0$, which splits into three peaks as $|\a|$ and $|\beta|$ increase; this solution is called a ``rogue wave triplet" in \citep{refAKA,refKAA11} and the ``three sisters'' in \citep{refGail11}.

\section{\label{sec3}The Boussinesq equation}
\subsection{Introduction}
Clarkson and Kruskal \citep{refCK} showed that Boussinesq equation \eqref{eq:bq} has symmetry reductions to 
the first, second and fourth \p\ equations\ (\PI, \PII, \PIV)  
\begin{align}\label{eq:PI} 
w'' &= 6w^2 + z , \\
\label{eq:PII} 
w'' &= 2w^3 + z w + \a, \\ 
\label{eq:PIV} 
w'' &= 
\frac{(w')^2}{2w} + \frac{3}{2}w^3 + 4z w^2 + 2(z^2 - \a)w + \frac{\beta}{w},
\end{align}
with $'=\d\/\d z$, and $\alpha$, $\beta$ arbitrary constants.
Vorob'ev \citep{refVor} and Yablonskii \citep{refYab59} expressed the rational solutions of \PII\ \eqref{eq:PII} in terms of polynomials, now known as the \emph{Yablonskii--Vorob'ev polynomials} (see also \citep{refCM03}). Okamoto \citep{refOkamotoiii} derived analogous polynomials, now known as the \emph{Okamoto polynomials}, related to some of the rational solutions of \PIV\ \eqref{eq:PIV}. Subsequently Okamoto's results were generalized by Noumi and Yamada \citep{refNY99i} who showed that all rational solutions of \PIV\ can be expressed in terms of logarithmic derivatives of two sets of special polynomials, called the \emph{generalized Hermite polynomials} and the \emph{generalized Okamoto polynomials} (see also \citep{refPAC03piv}). Consequently rational solutions of \eqref{eq:bq} can be obtained in terms the Yablonskii--Vorob'ev, generalized Hermite and generalized Okamoto polynomials, cf.~\citep{refPAC08}. 
Some of the rational solutions that are expressed in terms of the generalized Okamoto polynomials are generalized to give the rational solutions of the Boussinesq equation \eqref{eq:bq} obtained in \citep{refPAC08,refGPS,refPel98}, which are analogs of the rational solutions of the KdV equation \eqref{eq:kdv} \citep{refAbSat,refAdMoser,refAMM77,refCC77}.
However none of these rational solutions of the Boussinesq equation \eqref{eq:bq} are bounded for all real $x$ and $t$, so are unlikely to have any physical significance.

It is known that there are additional rational solutions of the Boussinesq equation \eqref{eq:bq} which don't arise from the above construction. For example, Ablowitz and Satsuma \citep{refAbSat} derived the rational solution
\begin{equation}\label{bq:rat1}
u(x,t)= 2\pderiv[2]{}{x}\ln(1+x^2+t^2)=\frac{4(1-x^2+t^2)}{(1+x^2+t^2)^2} ,
\end{equation} 
by taking a long-wave limit of the two-soliton solution, see also \citep{refTajMur91,refTajWat}. This solution is bounded for real $x$ and $t$, and tends to zero algebraically as $|x|,|t|\to\infty$. 

If in the Boussinesq equation \eqref{eq:bq}, we make the transformation \begin{equation}u(x,t)=2\pderiv[2]{}{x}\ln F(x,t),\end{equation}
then we obtain the bilinear equation
\begin{equation}FF_{tt}-F_t^2 + FF_{xx}-F_x^2 - \tfrac13\left(FF_{xxxx} -4 F_xF_{xxx}+3F_{xx}^2\right)=0,\label{eq:bilin}
\end{equation}
first derived by Hirota \cite{refHir73},
which can be written in the form
\begin{equation}(\D_t^2 + \D_x^2-\tfrac13\D_x^4)F\cdot F=0,\end{equation}
where $\D_x$ and $\D_t$ are Hirota operators \eqref{hirotaop}.

\subsection{Rational solutions of the Boussinesq equation}
Since the Boussinesq equation \eqref{eq:bq} has the rational solution \eqref{bq:rat1} then we seek solutions in the form
\begin{equation}\label{rat:ansatz1}
{u}_n(x,t)  = 2\pderiv[2]{}{x}\ln {F}_n(x,t),\qquad n\geq1,\end{equation}
where $F_{n}(x,t)$ is a polynomial of degree $\tfrac12n(n+1)$ in $x^2$ and $t^2$, with total degree $\tfrac12n(n+1)$, of the form
\begin{equation}\label{rat:ansatz2}
{F}_n(x,t) = \sum_{m=0}^{n(n+1)/2}\sum_{j=0}^ma_{j,m}x^{2j}t^{2(m-j)},
\end{equation}
with  $a_{j,m}$ constants which are determined by equating powers of $x$ and $t$. Using this procedure we obtain the following polynomials 
\begin{subequations}\label{bq:fn}\begin{align} 
F_1(x,t)&=x^2+t^2+1,\\
F_2(x,t)&={x}^{6}+ \left( 3{t}^{2}+\tfrac{25}3 \right) {x}^{4}+ \left( 3{t}^{4}+30{t}^{2}-\tfrac{125}9 \right) {x}^{2}+{t}^{6}+\tfrac{17}3{t}^{4}+\tfrac{475}9{t}^{2}+\tfrac{625}9,\\
F_3(x,t)&={x}^{12}+ \left( 6{t}^{2}+\tfrac{98}3 \right) {x}^{10}+ \left( 15{t}^{4}+230{t}^{2}+\tfrac{245}3 \right) {x}^{8}
+ \left( 20{t}^{6}+\tfrac{1540}3{t}^{4}+\tfrac{18620}9{t}^{2}+{\tfrac {75460}{81}} \right) {x}^{6}\nonumber\\
&\phantom{={x}^{12}\ }+ \left( 15{t}^{8}+\tfrac{1460}3{t}^{6}+\tfrac{37450}9{t}^{4}+\tfrac{24500}3{t}^{2}-{\tfrac {5187875}{243}}\right) {x}^{4} \nonumber\\
&\phantom{={x}^{12}\ }+ \left( 6{t}^{10}+190{t}^{8}+\tfrac{35420}9{t}^{6}-\tfrac{4900}9{t}^{4}+\tfrac{188650}{27}{t}^{2}+{\tfrac {159786550}{729}} \right) {x}^{2}\nonumber\\ 
&\phantom{={x}^{12}\ }+{t}^{12}+\tfrac{58}3{t}^{10}+\tfrac{1445}3{t}^{8}+{\tfrac {798980}{81}}{t}^{6}+{\tfrac {16391725}{243}}{t}^{4}
+{\tfrac {300896750}{729}}{t}^{2}+{\tfrac {878826025}{6561}},
\end{align}\end{subequations}
and the polynomials $F_4(x,t)$ and $F_5(x,t)$ are given in the Appendix.
We note that these polynomials have the following form
\[ F_n(x,t)= \big(x^2+t^2\big)^{n(n+1)/2} + G_{n}(x,t),\]
where $G_n(x,t)$ is a polynomial of degree $\tfrac12(n+2)(n-1)$ in both $x^2$ and $t^2$. 
We remark that the polynomials $F_n(x,t)$ which arise in the rational solutions of the focusing NLS equation \eqref{eq:fnls} have a similar structure, see for example \eqref{nls:f2}, 
though the coefficients in the polynomials $G_n(x,t)$ are different. The polynomials $F_j(x,t)$, for $j=2,3,4$, in scaled variables, are given by Pelinovsky and Stepanyants \citep{refPelStep} -- see their equations (6)--(8). However whilst they state that the polynomials are associated with solutions of their equation (2), which is a scaled variant of the Boussinesq equation \eqref{eq:bq}, Pelinovsky and Stepanyants don't mention, or reference, the Boussinesq equation.

\begin{figure}[ht]
\[\begin{array}{c@{\quad}c@{\quad}c} 
\includegraphics[width=2in]{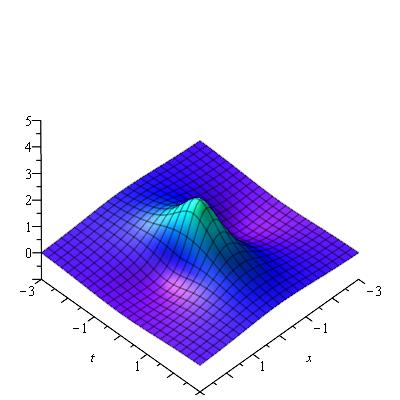} & \includegraphics[width=2in]{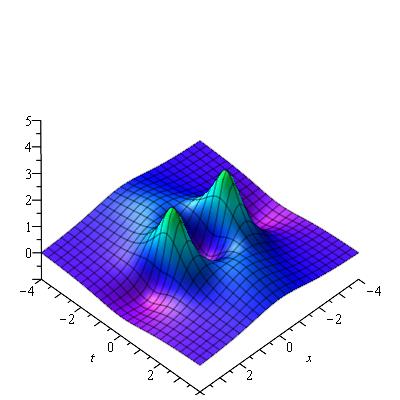}  &
\includegraphics[width=2in]{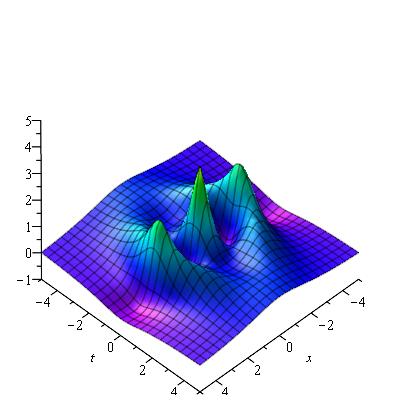} \\ u_1(x,t) & u_2(x,t) & u_3(x,t)\\ 
\includegraphics[width=2in]{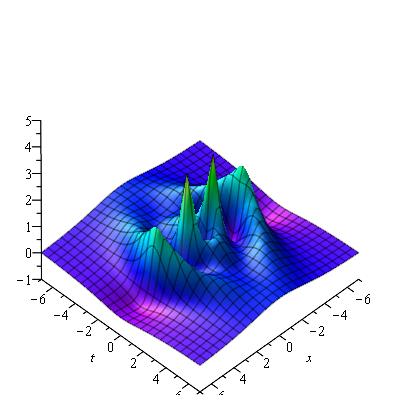} & \includegraphics[width=2in]{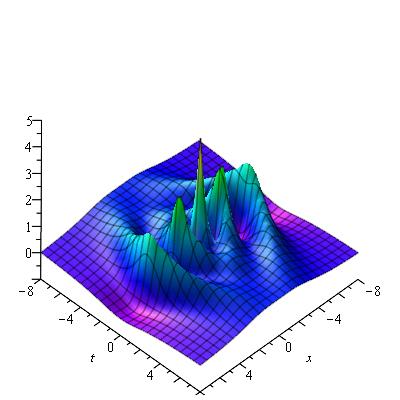} & 
\includegraphics[width=2in]{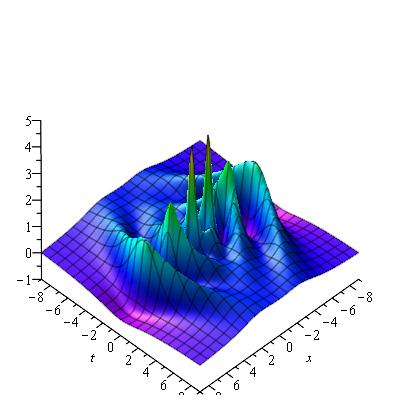} \\
u_4(x,t) & u_5(x,t) & u_6(x,t)
\end{array}\]
\caption{\label{fig31}Plots of the rational solutions $u_n(x,t)$, for $n=1,2,\ldots,6$, of the Boussinesq equation. }
\end{figure}%
In Figure \ref{fig31}, plots of the rational solutions $u_n(x,t)$, for $n=1,2,\ldots,6$, of the Boussinesq equation. These show that the maxima of the solutions all lie on the line $t=0$, with $n$ local maxima for the rational solution $u_n(x,t)$.

\begin{figure}[ht]
\[\begin{array}{c@{\quad}c@{\quad}c} 
\includegraphics[width=2in]{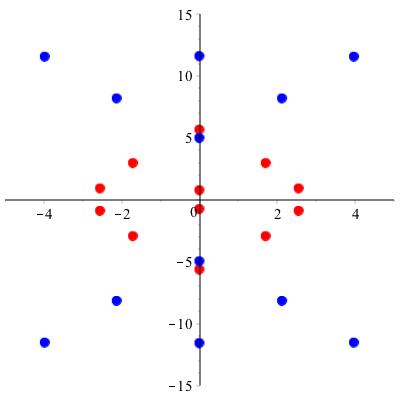} & \includegraphics[width=2in]{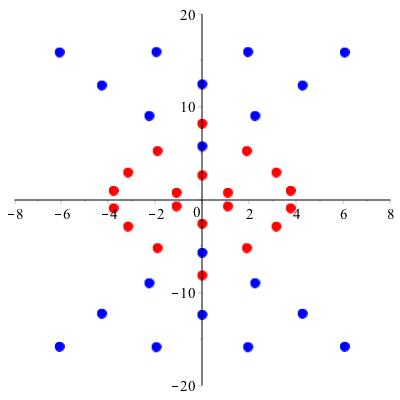} &\includegraphics[width=2in]{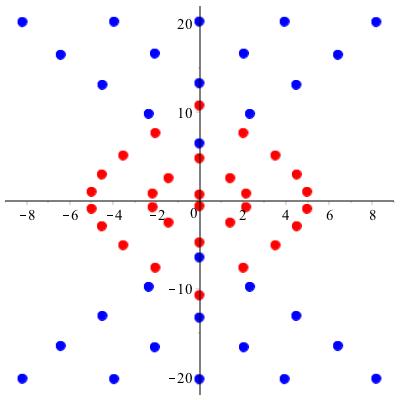}\\
F_3(x,t) & F_4(x,t) & F_5(x,t)
\end{array}\]
\caption{\label{fig32a}Plots of the complex roots of the polynomials $F_n(x,t)$, for $3,4,5$, for $t=0$ (red) and $t=3n$ (blue), i.e.\ $t=9$ for $n=3$, $t=12$ for $n=4$ and $t=15$ for $n=5$. Each plot shows the complex $x$-plane with roots in $x$ of $F_n(x,t)$ are shown at two different values of $t$.}
\end{figure}

\begin{figure}[ht]
\[\begin{array}{c@{\quad}c@{\quad}c} 
\includegraphics[width=2in]{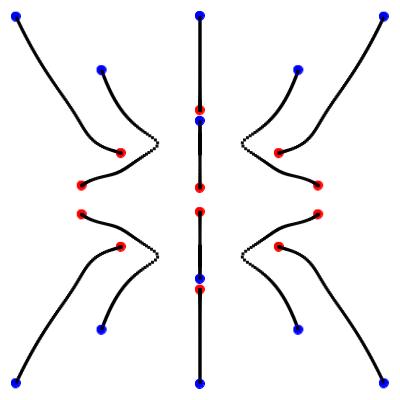} & \includegraphics[width=2in]{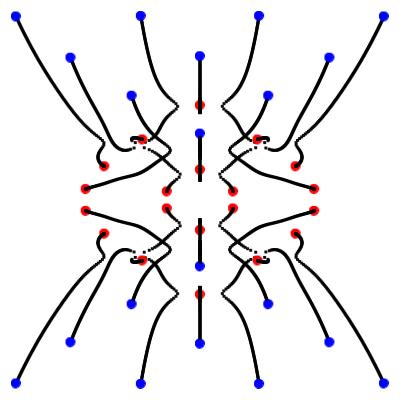} &\includegraphics[width=2in]{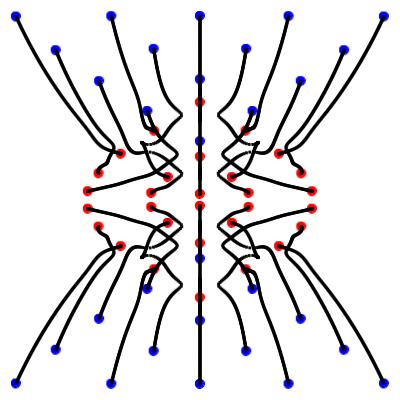}\\
F_3(x,t) & F_4(x,t) & F_5(x,t)
\end{array}\]
\caption{\label{fig32b}Plots of the loci of the complex roots of $F_n(x,t)$, for $3,4,5$, as $t$ varies, with $t=0$ (red) and $t=3n$ (blue), i.e.\ $t=9$ for $n=3$, $t=12$ for $n=4$ and $t=15$ for $n=5$.}
\end{figure}

\begin{figure}[ht]
\[\begin{array}{c@{\quad}c@{\quad}c} 
\includegraphics[width=2in]{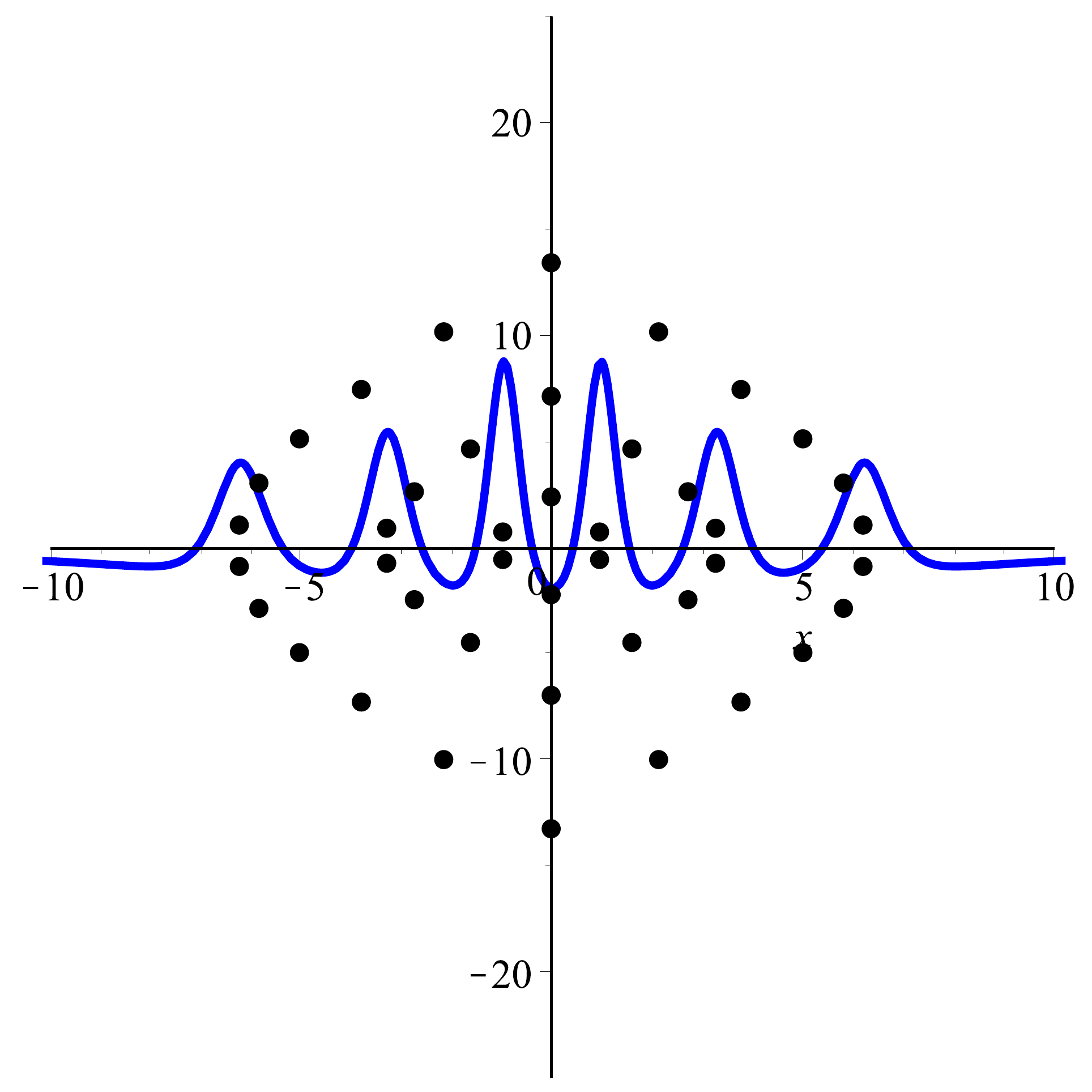}& \includegraphics[width=2in]{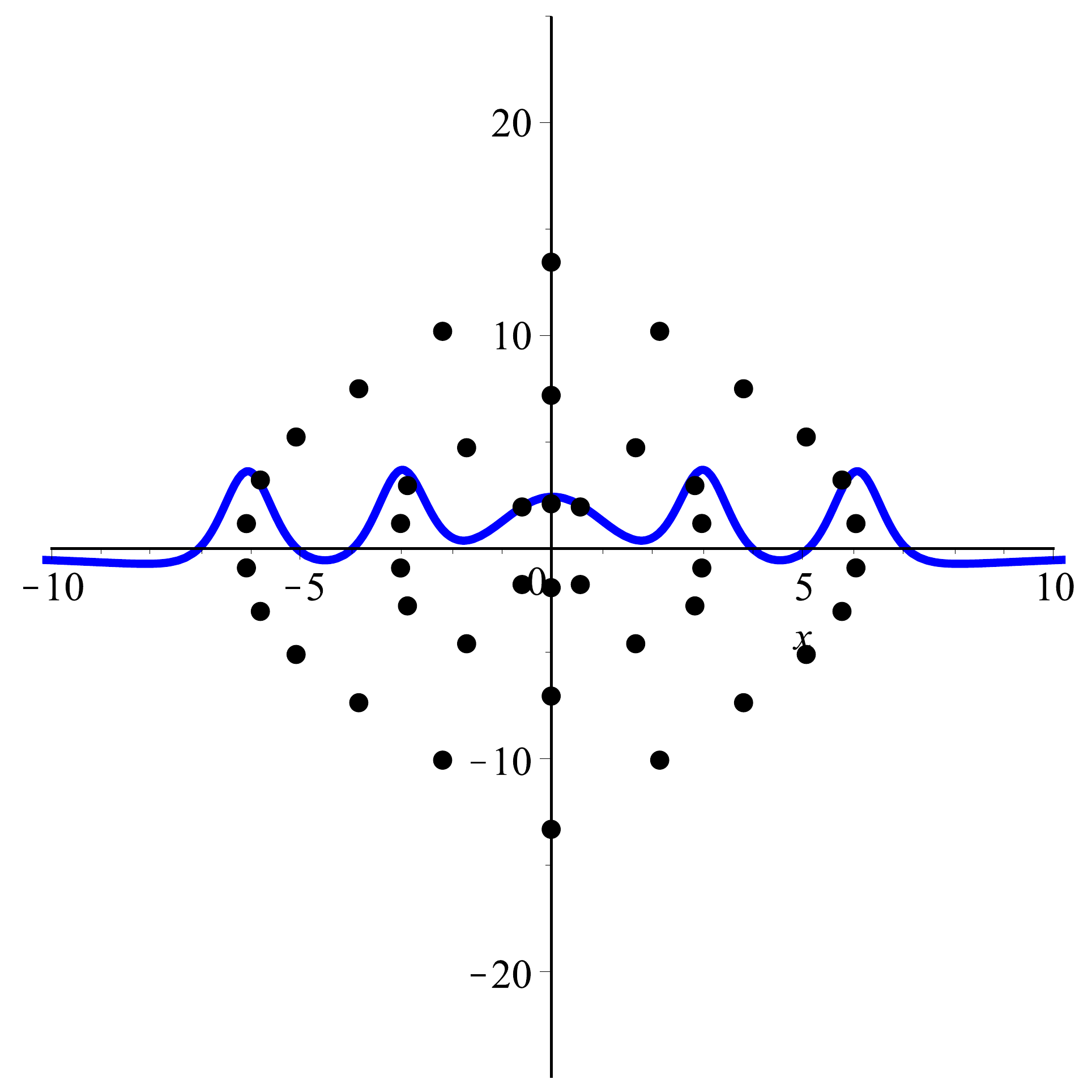} & \includegraphics[width=2in]{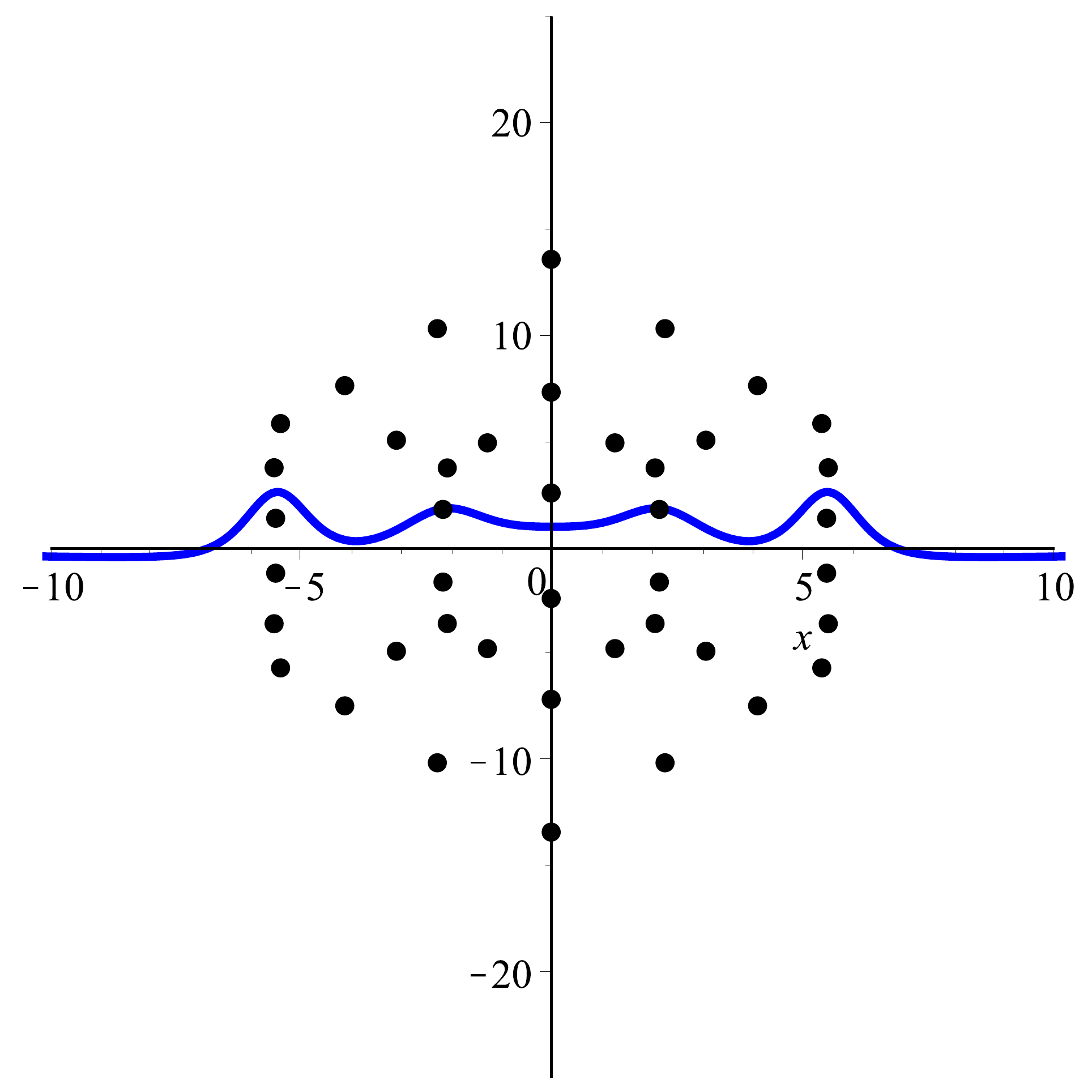} \\ t=0 & t=1 &t=2.5 \\
\includegraphics[width=2in]{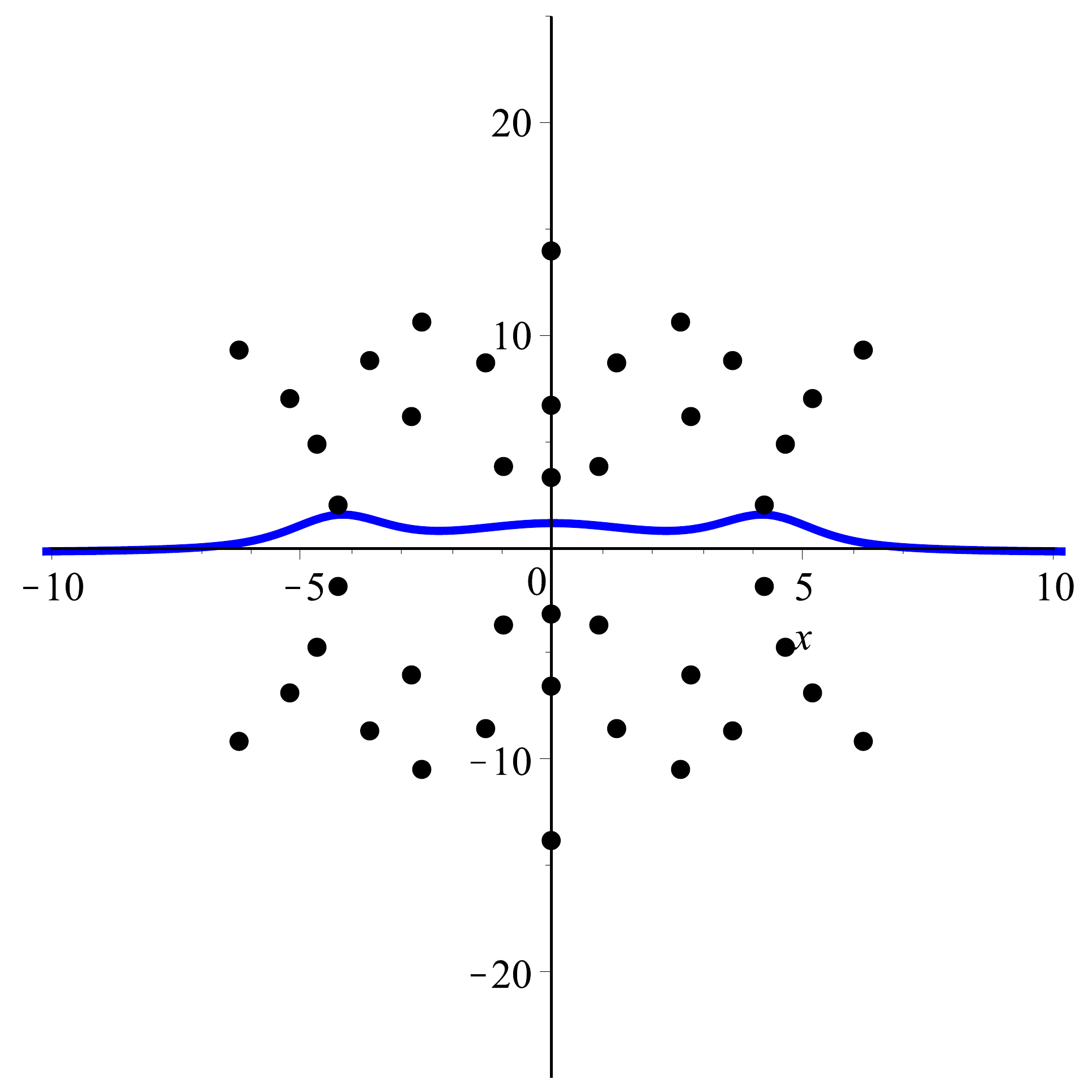} & \includegraphics[width=2in]{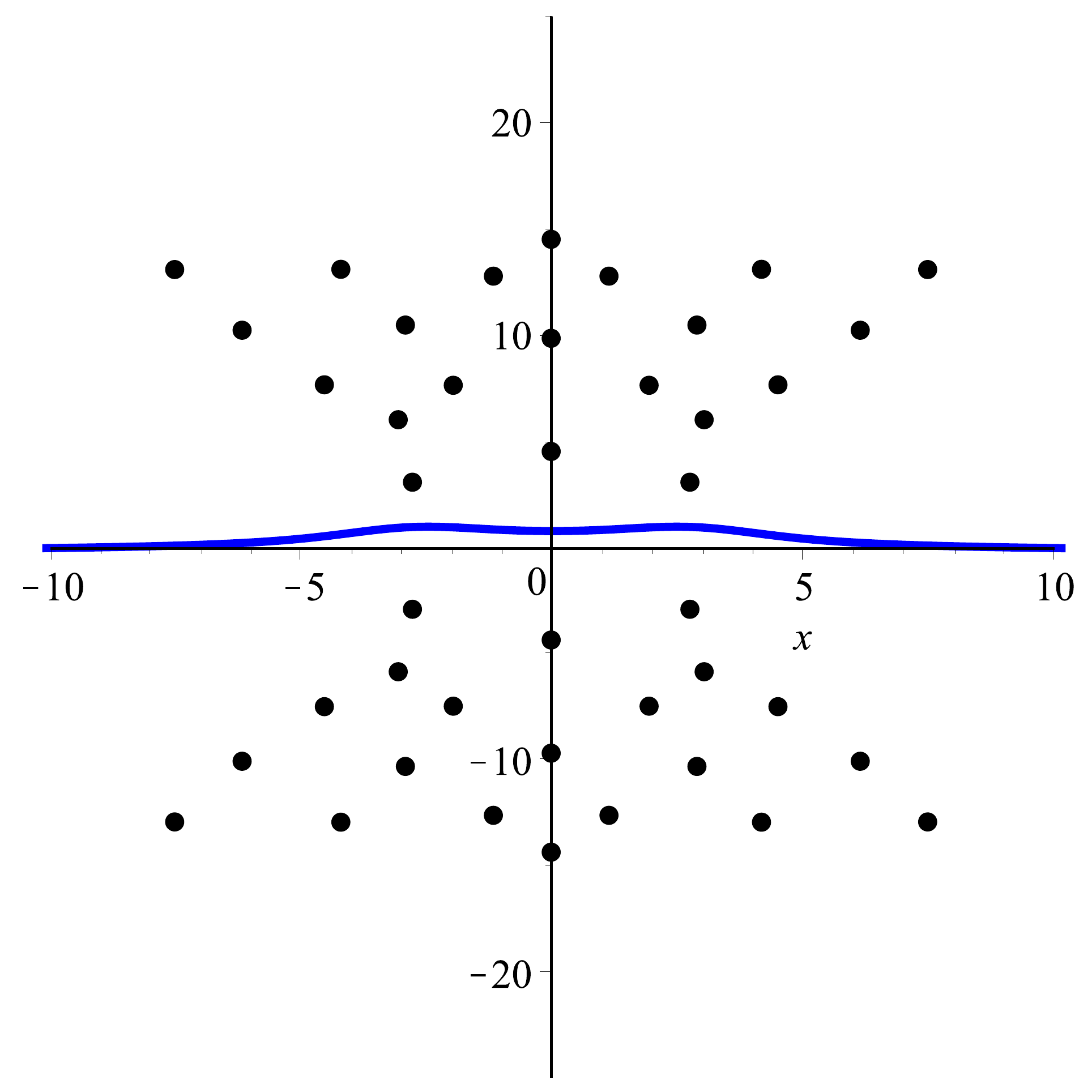} & \includegraphics[width=2in]{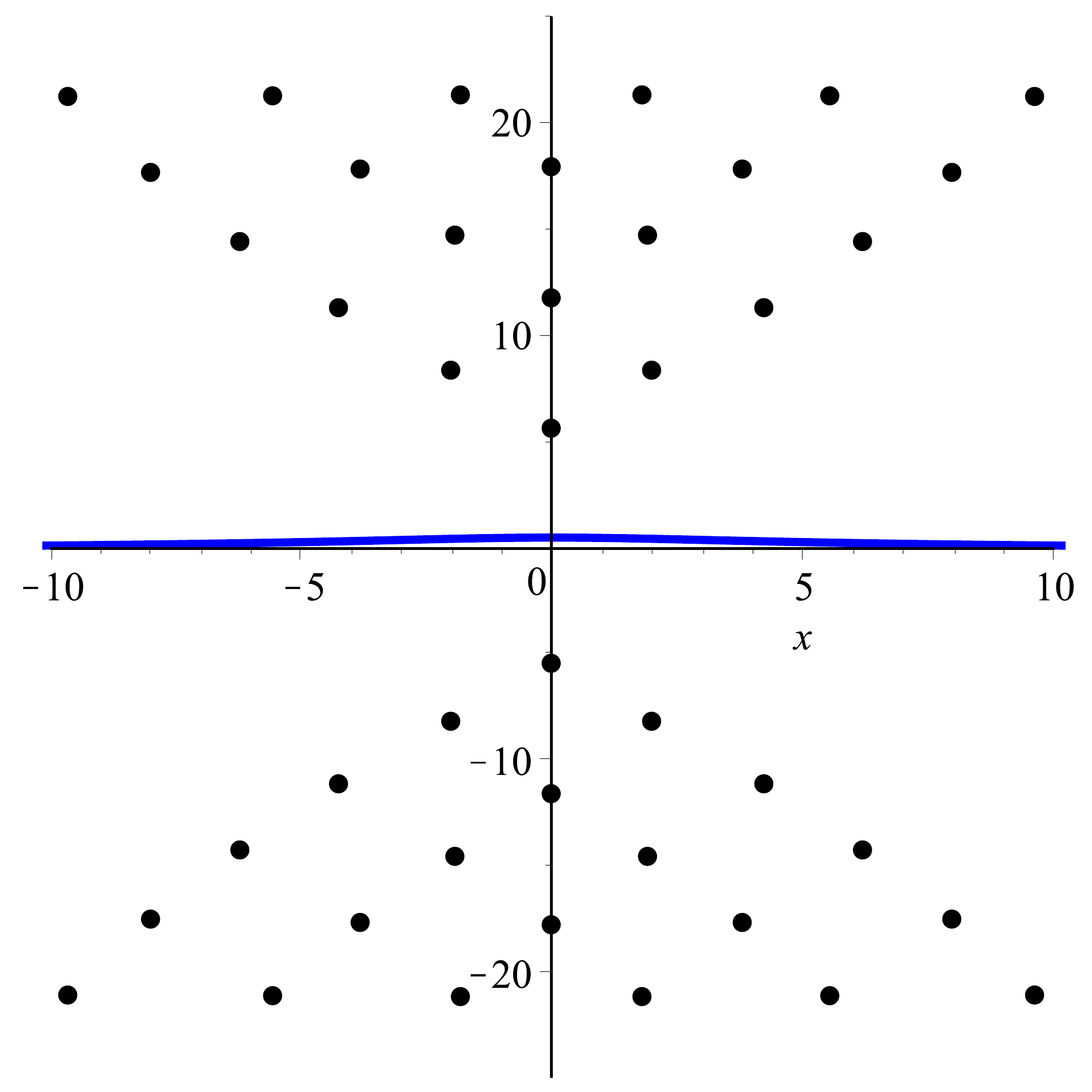} \\
t=5 & t=8 & t=15  
\end{array}\]
\caption{\label{fig34}Plots of the loci of the complex roots of $F_6(x,t)$ with the solution $u_6(x,t)$ superimposed (blue), as $t$ varies. The scale on the vertical axis relates to the complex $x$-plane for the roots of $F_6(x,t)$.}\end{figure}

In Figure \ref{fig32a}, plots of the complex roots of $F_n(x,t)$, for $3,4,5$, for $t=0$ and $t=3n$, i.e.\ $t=9$ for $n=3$, $t=12$ for $n=4$ and $t=15$ for $n=5$, are given. Each plot shows the complex $x$-plane with roots in $x$ of $F_n(x,t)$ are shown at two different values of $t$. These show a ``triangular" structure for both $t=0$ and $t=3n$, though with a different orientation. For $t=0$ the roots of the polynomials approximately form two isosceles triangles with curved sides. For $t=3n$ the roots of the polynomials again approximately form two isosceles triangles, though the values of the roots
show that they actually also lie on curves rather than straight lines. An analogous situation arises for the Yablonskii--Vorob'ev polynomials \cite{refCM03} and generalized Okamoto polynomials \cite{refPAC03piv}.

In Figure \ref{fig32b}, plots of the loci of the complex roots of $F_n(x,t)$, for $3,4,5$, as $t$ varies, between between the ``triangular" structures for $t=0$ and $t=3n$ are given. These show that as $t$ increases the roots move away from the real axis.

In Figure \ref{fig34}, plots of the loci of the complex roots of $F_6(x,t)$ with the solution $u_6(x,t)$ superimposed, as $t$ varies are given. The scale on the vertical axis relates to the complex $x$-plane for the roots of $F_6(x,t)$. These show that as the roots move away from the real axis, the solution decays to zero. 

\subsection{Generalised rational solutions of the Boussinesq equation}
Since the focusing NLS equation \eqref{eq:fnls} has generalised rational solutions, see \eqref{nls:u2hat}, then a natural question is whether 
the Boussinesq equation \eqref{eq:bq} also has generalised rational solutions. To investigate this, we are concerned with the following theorem.
\begin{theorem}\label{thm:Fntilde}
The Boussinesq equation \eqref{eq:bq} has generalised rational solutions  in the form
\begin{equation} \widetilde{u}_n(x,t;\alpha,\beta)  = 2\pderiv[2]{}{x}\ln \widetilde{F}_n(x,t;\alpha,\beta),\label{untilde}\end{equation}
for $n\geq1$, with
\begin{equation}\label{bq:untilde}
\widetilde{F}_{n+1}(x,t;\alpha,\beta)= F_{n+1}(x,t) +2\a tP_{n}(x,t)+2\b xQ_{n}(x,t)+\big(\a^2+\beta^2\big)F_{n-1}(x,t),\end{equation}
where $F_n(x,t)$ is given by \eqref{bq:fn}, 
$P_{n}(x,t)$ and $Q_{n}(x,t)$ are polynomials of degree $\tfrac12n(n+1)$ in $x^2$ and $t^2$, and $\alpha$ and $\beta$ are arbitrary constants.
\end{theorem}
Since the generalised polynomial $\widehat{D}_2(x,t;\alpha,\beta)$ for the focusing NLS equation has the structure given by \eqref{nls:F2hat}, we suppose that the Boussinesq equation \eqref{eq:bq} has a solution in the form \eqref{untilde}, with $F_n(x,t)$  given by \eqref{bq:fn} and
the polynomials $P_{n}(x,t)$ and $Q_{n}(x,t)$, which are of degree $\tfrac12n(n+1)$ in $x^2$ and $t^2$, have the form
\begin{align}\label{bq:PQform}
P_n(x,t) &= \sum_{m=0}^{n(n+1)/2}\sum_{j=0}^m b_{j,m}x^{2j}t^{2(m-j)},\qquad
Q_n(x,t) = \sum_{m=0}^{n(n+1)/2}\sum_{j=0}^m c_{j,m}x^{2j}t^{2(m-j)},
\end{align}
where the coefficients $b_{j,m}$ and $c_{j,m}$ are to be determined. 
Substituting \eqref{bq:untilde} into the bilinear equation \eqref{eq:bilin} with
$F_1(x,t)$, $F_2(x,t)$, $F_3(x,t)$ and $F_4(x,t)$ given by \eqref{bq:fn}, $P_{n}(x,t)$ and $Q_{n}(x,t)$ in the form \eqref{bq:PQform}, then by equating powers of $x$ and $t$ we find that
\begin{subequations}\label{bq:PQ}\begin{align}
P_1(x,t)&=3x^2-t^2+\tfrac53,\\
Q_1(x,t)&=x^2-3 t^2-\tfrac13,\\
P_2(x,t)&=5x^6 -\left(5t^2-35\right)x^4-\left(9t^4+\tfrac{190}3t^2+\tfrac{665}{9}\right)x^2+t^6-\tfrac73t^4-\tfrac{245}9t^2+\tfrac{18865}{81},\\
Q_2(x,t)&=x^6-\left(9t^2-\tfrac{13}3\right)x^4-\left(5t^4+\tfrac{230}3t^2+\tfrac{245}9\right)x^2 +5t^6+15t^4+\tfrac{535}9t^2+\tfrac{12005}{81},
\end{align}\end{subequations}
with $\alpha$ and $\beta$ arbitrary constants; the polynomials $P_3(x,t)$, $Q_3(x,t)$, $P_4(x,t)$ and $Q_4(x,t)$ are given in the Appendix.

\begin{figure}[ht]
\[\begin{array}{c@{\quad}c@{\quad}c} 
\includegraphics[width=2in]{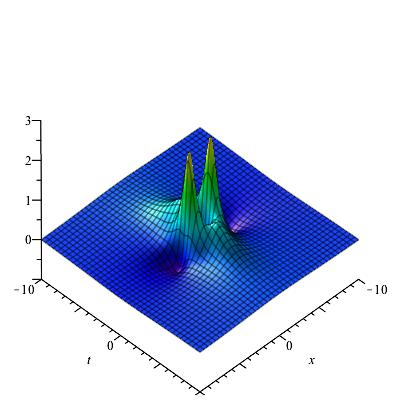}& \includegraphics[width=2in]{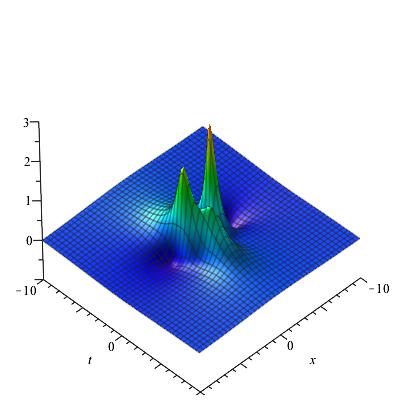}
& \includegraphics[width=2in]{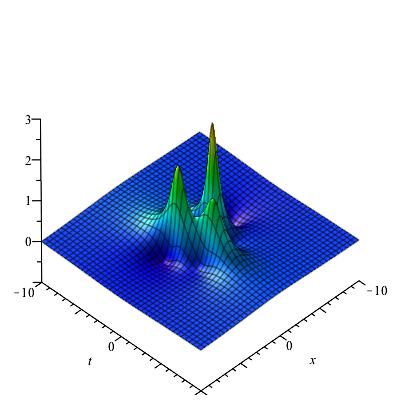}\\
\alpha=\beta=0 &  \alpha=\beta=5& \alpha=\beta=10 \\
\includegraphics[width=2in]{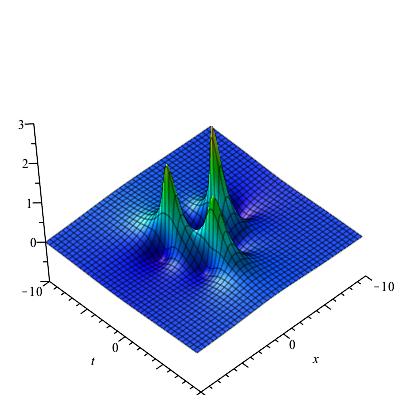} & \includegraphics[width=2in]{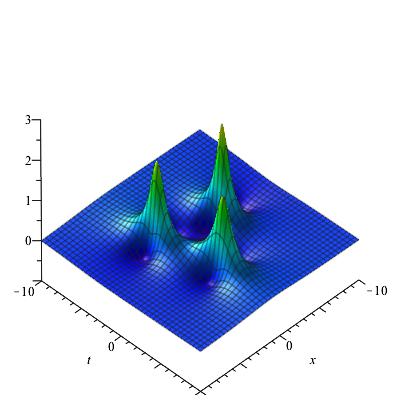}& \includegraphics[width=2in]{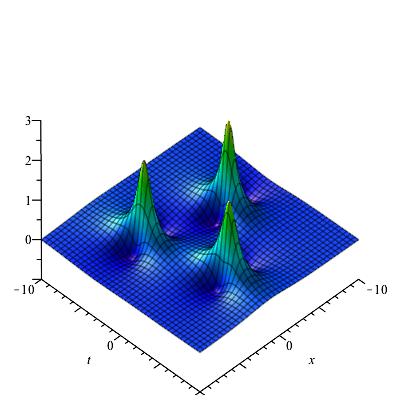}\\
\alpha=\beta=20& \alpha=\beta=50  & \alpha=\beta=100 
\end{array}\]
\caption{\label{fig:u2gen}Plots of the generalised rational solution $\widetilde{u}_2(x,t;\alpha,\beta)$ of the Boussinesq equation for various values of the parameters $\alpha$ and $\beta$.}
\end{figure}

\begin{figure}[ht]
\[\begin{array}{c@{\quad}c@{\quad}c} 
\includegraphics[width=2in]{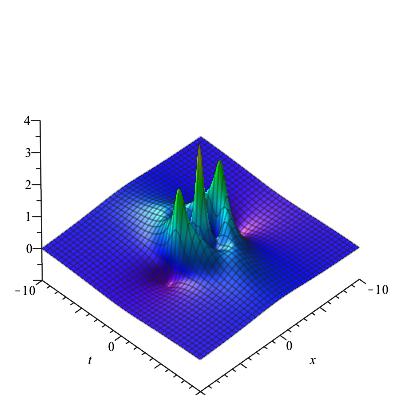}& \includegraphics[width=2in]{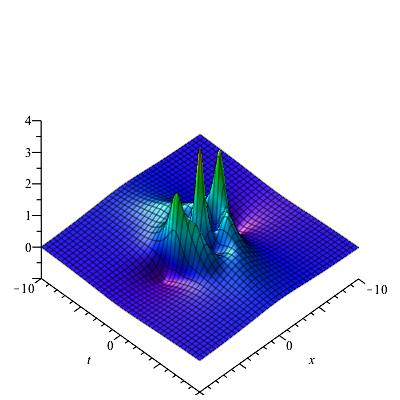}
& \includegraphics[width=2in]{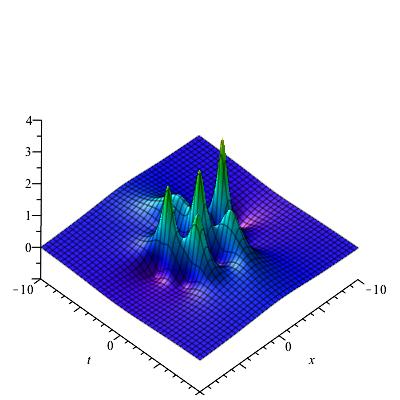} \\
\alpha=\beta=0 &  \alpha=\beta=100 & \alpha=\beta=500  \\
\includegraphics[width=2in]{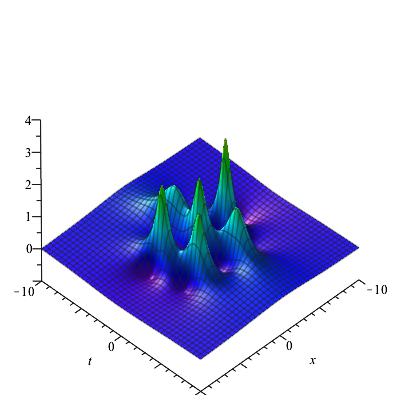}&  \includegraphics[width=2in]{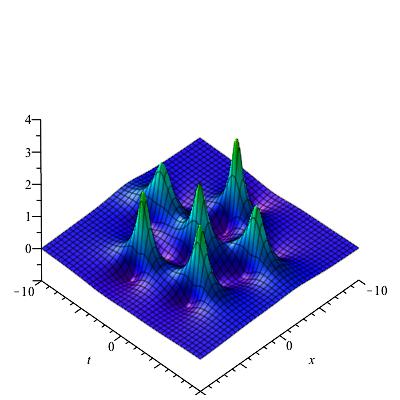}& \includegraphics[width=2in]{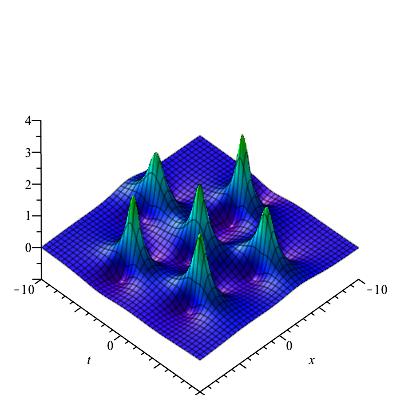}\\
 \alpha=\beta=1000 &  \alpha=\beta=5000  & \alpha=\beta=10000 
\end{array}\]
\caption{\label{fig:u3gen}Plots of the generalised rational solution $\widetilde{u}_3(x,t;\alpha,\beta)$ of the Boussinesq equation for various values of the parameters $\alpha$ and $\beta$.}
\end{figure}

\begin{figure}[ht]
\[\begin{array}{c@{\quad}c@{\quad}c} 
\includegraphics[width=2in]{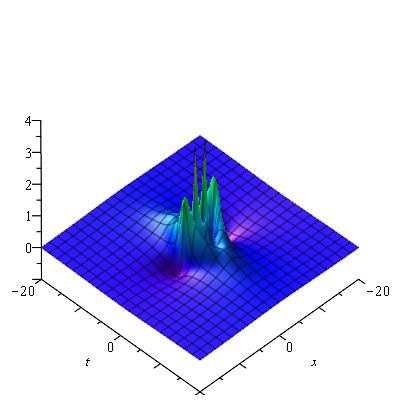}& \includegraphics[width=2in]{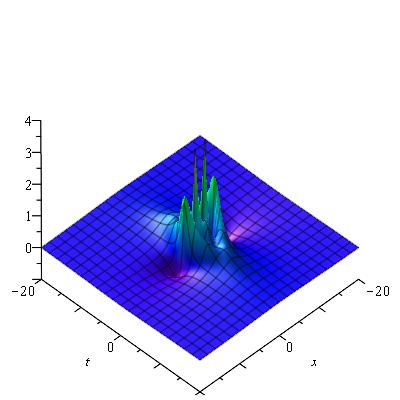}
& \includegraphics[width=2in]{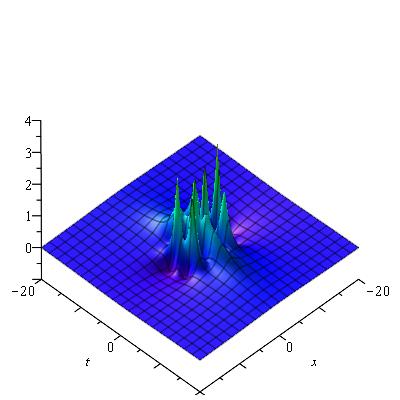} \\
\alpha=\beta=0 &  \alpha=\beta=10^3 & \alpha=\beta=10^5  \\
\includegraphics[width=2in]{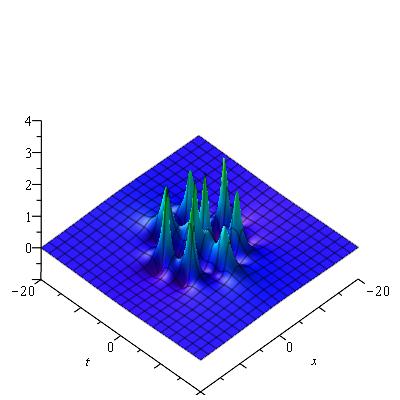}&  \includegraphics[width=2in]{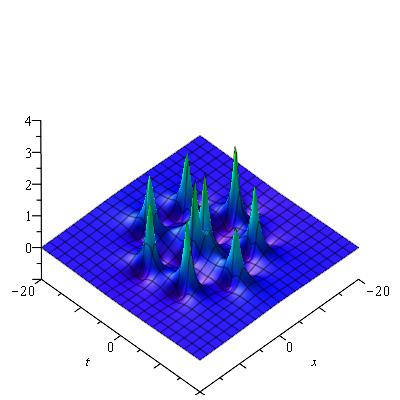}& \includegraphics[width=2in]{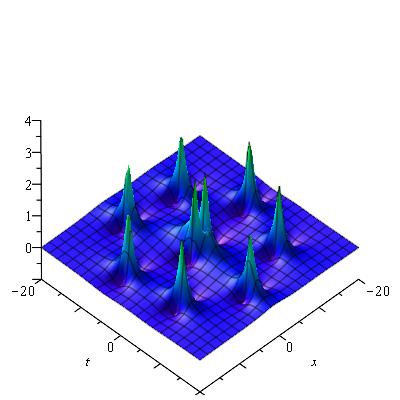}\\
 \alpha=\beta=10^6 &  \alpha=\beta=10^7  & \alpha=\beta=10^8
\end{array}\]
\caption{\label{fig:u4gen}Plots of the generalised rational solution $\widetilde{u}_4(x,t;\alpha,\beta)$ of the Boussinesq equation for various values of the parameters $\alpha$ and $\beta$.}
\end{figure}

\begin{figure}[ht]
\[\begin{array}{c@{\quad}c@{\quad}c} 
\includegraphics[width=2in]{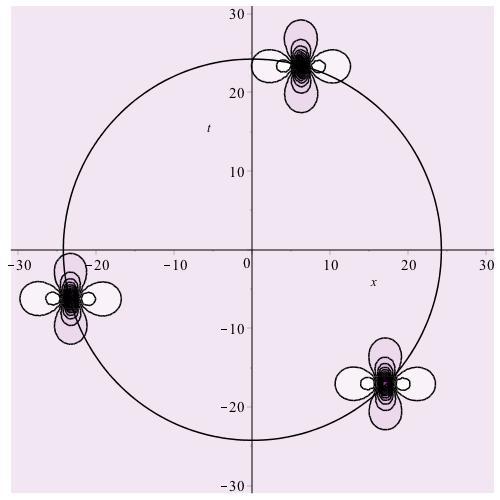}& \includegraphics[width=2in]{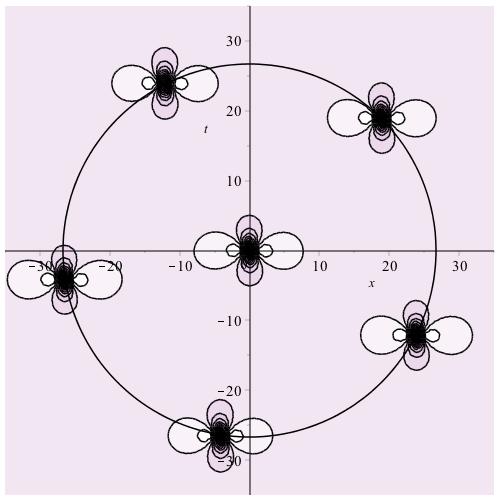}
& \includegraphics[width=2in]{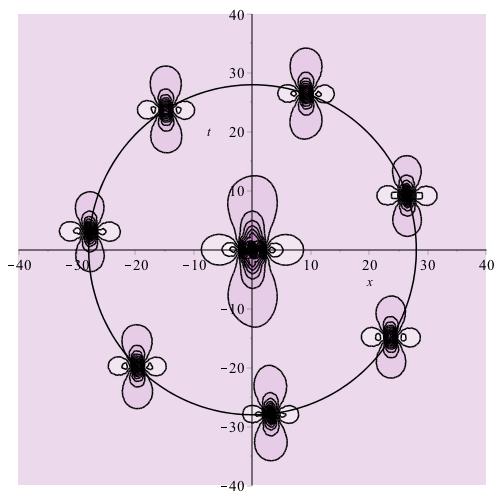} \\
\widetilde{u}_2(x,t;10^4,10^4) & \widetilde{u}_3(x,t;10^7,10^7)  & \widetilde{u}_4(x,t;10^{10},10^{10})
\end{array}\]
\caption{\label{fig:u234cont}Contour plots of the generalised rational solutions $\widetilde{u}_2(x,t;10^4,10^4)$, $\widetilde{u}_3(x,t;10^7,10^7)$ and $\widetilde{u}_4(x,t;10^{10},10^{10})$ of the Boussinesq equation.}
\end{figure}

The first two generalised rational solutions are
\begin{align}\label{bq:genrat2} \widetilde{u}_2(x,t;\alpha,\beta)  &=  2\pderiv[2]{}{x}\ln \widetilde{F}_2(x,t;\alpha,\beta),\\
\label{bq:genrat3}\widetilde{u}_3(x,t;\alpha,\beta)  &= 2\pderiv[2]{}{x}\ln \widetilde{F}_3(x,t;\alpha,\beta),
\end{align}
where
\begin{align}\widetilde{F}_2(x,t;\alpha,\beta) &=F_2(x,t) + {2\a t P_{1}(x,t)}+{2\b x Q_{1}(x,t)} +{\a^2}+{\beta^2}\nonumber\\
&={x}^{6}+ \left( 3{t}^{2}+\tfrac{25}3 \right) {x}^{4}+ \left( 3{t}^{4}+30{t}^{2}-\tfrac{125}9 \right) {x}^{2}+{t}^{6}+\tfrac{17}3{t}^{4}+\tfrac{475}9{t}^{2}+\tfrac{625}9\nonumber\\
&\phantom{=x^6\ }+ {2\a t\left(3x^2-t^2+\tfrac53\right)} + {2\b x \left(x^2-3 t^2-\tfrac13\right)}+{\a^2}+{\beta^2},\end{align}
and \begin{align}
\widetilde{F}_3(x,t;\alpha,\beta)&={F}_3(x,t)+ {2\a t P_{2}(x,t)}+{2\b x Q_{2}(x,t)} +({\a^2}+{\beta^2})F_{1}(x,t)\nonumber\\
&={x}^{12}+\left( 6{t}^{2}+\tfrac{98}3 \right) {x}^{10}+ \left( 15{t}^{4}+230{t}^{2}+\tfrac{245}3 \right) {x}^{8} 
+ \left( 20{t}^{6}+\tfrac{1540}3{t}^{4}+\tfrac{18620}9{t}^{2}+{\tfrac {75460}{81}} \right) {x}^{6}\nonumber\\
&\phantom{={x}^{12}\ }+ \left( 15{t}^{8}+\tfrac{1460}3{t}^{6}+\tfrac{37450}9{t}^{4}+\tfrac{24500}3{t}^{2}-{\tfrac {5187875}{243}}\right) {x}^{4} \nonumber\\
&\phantom{={x}^{12}\ }+ \left( 6{t}^{10}+190{t}^{8}+\tfrac{35420}9{t}^{6}-\tfrac{4900}9{t}^{4}+\tfrac{188650}{27}{t}^{2}+{\tfrac {159786550}{729}} \right) {x}^{2}\nonumber\\
&\phantom{={x}^{12}\ }+{t}^{12}+\tfrac{58}3{t}^{10}+\tfrac{1445}3{t}^{8}+{\tfrac {798980}{81}}{t}^{6}+{\tfrac {16391725}{243}}{t}^{4}
+{\tfrac {300896750}{729}}{t}^{2}+{\tfrac {878826025}{6561}}\nonumber\\
&\phantom{={x}^{12}\ }+ {{2\a t\left\{5x^6 -\left(5t^2-35\right)x^4-\left(9t^4+\tfrac{190}3t^2+\tfrac{665}{9}\right)x^2+t^6-\tfrac73t^4-\tfrac{245}9t^2+\tfrac{18865}{81}\right\}}}\nonumber\\
&\phantom{={x}^{12}\ }+ {{2\b x\left\{x^6-\left(9t^2-\tfrac{13}3\right)x^4-\left(5t^4+\tfrac{230}3t^2+\tfrac{245}9\right)x^2 +5t^6+15t^4+\tfrac{535}9t^2+\tfrac{12005}{81}\right\}}}\nonumber\\
&\phantom{={x}^{12}\ }+ ({\a^2}+{\beta^2})(x^2+t^2+1),
\end{align}
with $\alpha$ and $\beta$ arbitrary constants.
Plots of the solutions $\widetilde{u}_2(x,t;\alpha,\beta)$, $\widetilde{u}_3(x,t;\alpha,\beta)$ and $\widetilde{u}_4(x,t;\alpha,\beta)$ of the Boussinesq equation for various values of the parameters $\alpha$ and $\beta$ are given in Figures \ref{fig:u2gen}, \ref{fig:u3gen} and \ref{fig:u4gen}, respectively. Contour plots of the solutions $\widetilde{u}_2(x,t;10^4,10^4)$, $\widetilde{u}_3(x,t;10^7,10^7)$ and $\widetilde{u}_4(x,t;10^{10},10^{10})$ of the Boussinesq equation \eqref{eq:bq} illustrating this behaviour are given in Figure \ref{fig:u234cont}. 

Figure \ref{fig:u2gen} shows that the solution $\widetilde{u}_2(x,t;\alpha,\beta)$ has two peaks when $\alpha=\beta=0$, then as $|\a|$ and $|\beta|$ increase a third peak appears. Numerical evidence suggests that as $|\a|$ and $|\beta|$ increase then the three peaks all tend to the same height $\max(\widetilde{u}_2)=4$. For $|\a|$ and $|\beta|$ sufficiently large, then $\widetilde{u}_2(x,t;\alpha,\beta)$ has three lumps which are essentially copies of the lowest-order solution, i.e.\ $u_1(x,t)$, which equally spaced on a circle; an analogous situation arises for the second generalised rational solution of the NLS equation \cite{refKAA12,refKAA13}.

Figure \ref{fig:u3gen} shows that the solution $\widetilde{u}_3(x,t;\alpha,\beta)$ has three peaks when $\alpha=\beta=0$, then as $|\a|$ and $|\beta|$ increase three more peaks appear, for $\alpha$ and $\beta$ sufficiently large with one central peak and five in a circle around it, so forming a pentagram. Again, numerical evidence suggests that as $|\a|$ and $|\beta|$ increase then the three peaks all tend to the same height $\max(\widetilde{u}_3)=4$. For $\alpha$ and $\beta$ sufficiently large, the rational solution $\widetilde{u}_3(x,t;\alpha,\beta)$ has six lumps, again essentially copies of the lowest-order solution $u_1(x,t)$, with five equally spaced on a circle; an analogous situation arises for the third generalised rational solution of the NLS equation \cite{refKAA11,refKAA13}.

Figure \ref{fig:u4gen} shows that the solution $\widetilde{u}_4(x,t;\alpha,\beta)$ has four peaks when $\alpha=\beta=0$, then as $|\a|$ and $|\beta|$ increase five more peaks appear, with for $\alpha$ and $\beta$ sufficiently large with two central peaks and seven in a ring around it, so forming a heptagram. As for $\widetilde{u}_2(x,t;\alpha,\beta)$ and $\widetilde{u}_3(x,t;\alpha,\beta)$, numerical evidence suggests that as $|\a|$ and $|\beta|$ increase then the peaks all tend to the same height $\max(\widetilde{u}_4)=4$.  An analogous situation arises for the fourth generalised rational solution of the NLS equation \cite{refKAA13}.

\begin{remark}{\rm  Ohta and Yang \citep[Figure 1]{refOY} show that for focusing NLS equation \eqref{eq:fnls}, the generalised rational solution $\widehat{\psi}_2(x,t;\alpha,\beta)$ \eqref{nls:u2hat} has a single peak when $\alpha=\beta=0$, and three peaks otherwise. 
Ohta and Yang \citep[Figure 2]{refOY} also show that the generalised rational solution $\widehat{\psi}_3(x,t;\alpha,\beta)$ has a single peak when unperturbed, and six peaks otherwise.
}\end{remark}

Define the polynomials $\Theta_n^{\pm} (x,t)$, for $n\in\mathbb{N}$, by 
\begin{equation}\label{bq:theta} \Theta_n^{\pm} (x,t)= xP_n(x,t)\pm\i t Q_n(x,t),
\end{equation} 
with $P_n(x,t)$ and $Q_n(x,t)$ the polynomials in the generalised rational solution \eqref{bq:untilde}. Then for $P_n(x,t)$ and $Q_n(x,t)$ given by \eqref{bq:PQ}, it is easily verified that $\Theta_n^{\pm} (x,t)$, for $n=1,2,3,4$, satisfy the bilinear equation \eqref{eq:bilin}. 
Hence in the general case we have the following conjecture.

\begin{conjecture}\label{conj:theta}
The polynomials $\Theta_n^{\pm} (x,t)$ given by \eqref{bq:theta}
satisfy the bilinear equation \eqref{eq:bilin}.
\end{conjecture}

Consequently, from this and Theorem \ref{bq:untilde} we have the following result.
\begin{lemma}Let $\Theta_n^{\pm} (x,t)$ be given by \eqref{bq:theta}, then the polynomial $\widetilde{F}_{n+1}(x,t;\alpha,\beta)$ given by \eqref{bq:untilde} can be written as
\begin{equation}\label{bq:untilde1}
\widetilde{F}_{n+1}(x,t;\alpha,\beta)= F_{n+1}(x,t) +(\a+\i\b)\Theta^+_{n}(x,t)+(\a-\i\b)\Theta^-_{n}(x,t)+\big(\a^2+\beta^2\big)F_{n-1}(x,t),\end{equation}
which is a linear combination of four solutions $F_{n+1}(x,t)$, $\Theta^{\pm}_{n}(x,t)$ and $F_{n-1}(x,t)$ of the bilinear equation \eqref{eq:bilin}.
\end{lemma}

\section{\label{sec4}Rational solutions of the \kp\ I equation}
\subsection{Introduction}
The \kp\ (KP) equation
\begin{equation}\label{eq:kp}
(v_\tau+6vv_\xi+v_{\xi\xi\xi})_\xi+3\sigma^2 v_{\eta\eta}=0, \qquad \sigma^2=\pm1,
\end{equation} 
which is known as KPI if $\sigma^2=-1$, i.e.\  \eqref{eq:kp1}, and KPII if $\sigma^2=1$,
was derived by Kadomtsev and Petviashvili \citep{refKP} to model ion-acoustic waves of small amplitude propagating in plasmas and is a two-dimensional generalisation of the KdV equation \eqref{eq:kdv}.
The KP equation arises in many physical applications including weakly two-dimen\-sion\-al long waves in shallow water \citep{refAS79,refSegurF}, where the sign of $\sigma^2$ depends upon the relevant magnitudes of gravity and surface tension, in nonlinear optics \citep{refPSK}, ion-acoustic waves in plasmas \citep{refInR}, two-dimensional matter-wave pulses in Bose-Einstein condensates \citep{refTDP}, and as a model for sound waves in ferromagnetic media \citep{refTF}.
The KP equation \eqref{eq:kp} is also a completely integrable soliton equation solvable by inverse scattering and again the sign of $\sigma^2$ is critical since if $\sigma^2=-1$, then the inverse scattering problem is formulated in terms of a Riemann-Hilbert problem \citep{refFA83,refManakov}, whereas for $\sigma^2=1$, it is formulated in terms of a $\overline\partial$ (``DBAR'') problem \citep{refABF}. 

The first rational solution of the KPI equation \eqref{eq:kp1},  is the so-called ``lump solution"
\begin{equation}\label{kp:lump}v(\xi,\eta,\tau)=2\pderiv[2]{}{\xi}\ln [(\xi-3\tau)^2+\eta^2+1] 
= -4\frac{(\xi-3\tau)^2-\eta^2-1}{[(\xi-3\tau)^2+\eta^2+1]^2},\end{equation}
which was found by Manakov \etal\ \citep{refMZBIM}. Subsequent studies of rational solutions of the KPI equation \eqref{eq:kp1} include Ablowitz \etal\ \citep{refACTV},  Ablowitz and Villarroel \citep{refAV,refVA99}, Dubard and Matveev \citep{refDubMat11,refDubMat13},
Gaillard \citep{refGail16a,refGail16b}, Johnson and Thompson \citep{refJT}, 
Ma \citep{refMa}, Pelinovsky \citep{refPel94,refPel98}, Pelinovsky and Stepanyants \citep{refPelStep}, 
Satsuma and Ablowitz \citep{refSatAb79}, and Singh and Stepanyants \citep{refSS}.

We remark that the KP equation \eqref{eq:kp} is invariant under the \textit{Galilean transformation}
\begin{equation}\label{eq:gt}
(\xi,\eta,\tau,v) \mapsto (\xi+6\la,\eta,\tau,v+\la),
\end{equation}
with $\la$ an arbitrary constant. In fact the rational solutions of the KPI equation \eqref{eq:kp1} derived by 
Dubard and Matveev \citep{refDubMat11,refDubMat13} and Gaillard \citep{refGail16a,refGail16b} are equivalent under the Galilean transformation \eqref{eq:gt}.

\subsection{Rational solutions of KPI related to the focusing NLS equation}
Dubard and Matveev \citep{refDubMat11,refDubMat13} derive rational solutions of the KPI equation \eqref{eq:kp1}
from the generalised rational solution $\widehat{\psi}_2(x,t;\alpha,\beta)$ \eqref{nls:u2hat} of the focusing NLS equation \eqref{eq:fnls}; see also \citep{refDGKM,refGail16a,refGail16b}. Specifically Dubard and Matveev \citep{refDubMat11,refDubMat13}
show that 
\begin{align}v(\xi,\eta,\tau)&
=2\pderiv[2]{}{\xi}\ln \widehat{D}_2(\xi-3\tau,\eta;\a,-48\tau) 
= \tfrac12\left(|\widehat{\psi}_2(x,t;\alpha,\beta)|^2-1\right)\Big|_{x=\xi-3\tau,t=\eta,\beta=-48\tau}, \label{kp1sol:nls}\end{align}
is a solution of the KPI equation \eqref{eq:kp1}.
If we define $F_2^{\rm nls}(\xi,\eta,\tau;\a)=\widehat{D}_2 (\xi-3\tau,\eta;\a,-48\tau)$, then 
\begin{align}
F_2^{\rm nls}(\xi,\tau;\a)
&={\xi}^{6}-18 \tau{\xi}^{5}+ 3\left(45 {\tau}^{2}+ {\eta}^{2}+{1}\right) {\xi}^{4} - 12\left(45 {\tau}^{2}+3 {\eta}^{2}{-\ 5} \right) \tau{\xi}^{3}\nonumber\\
&\phantom{={\xi}^{6}\ }+ \big\{ 3 {\eta}^{4}+ 18\left( 9 {\tau}^{2}{-\ 1} \right) {\eta}^{2}+1215 {\tau}^{4}{-\ 702 {\tau}^{2}+27} \big\} {\xi}^{2}
\nonumber\\
&\phantom{={\xi}^{6}\ }- \big\{ 18 \tau{\eta}^{4}+ 36\left(9 {\tau}^{2}+5  \right) \tau {\eta}^{2}+1458 {\tau}^{5}{-\ 2268 {\tau}^{3}+450 \tau} \big\}\xi \nonumber\\
&\phantom{={\xi}^{6}\ }+{\eta}^{6}+27 \left(  {\tau}^{2}+{1} \right) {\eta}^{4}+ 9\left( 27 {\tau}^{4}+{78 {\tau}^{2}+11} \right) {\eta}^{2}
+ 729 {\tau}^{6}{-\ 2349 {\tau}^{4}+3411 {\tau}^{2}+9}.\label{kp1F:nls2}
\end{align}
The polynomial $F_2^{\rm nls}(\xi,\tau;\a)$ satisfies
\begin{equation}\left(\D_\xi^4+\D_\xi\D_\tau-3\D_\eta^2\right)F_2\cdot F_2=0,\end{equation}
which is the bilinear form of the KPI equation \eqref{eq:kp1}, and so
\begin{equation}\label{kp1sol:nls2}
v_2^{\rm nls}(\xi,\eta,\tau;\a)=2\pderiv[2]{}{\xi}\ln F_2^{\rm nls}(\xi,\eta,\tau;\a),\end{equation}
is a rational solution of the KPI equation \eqref{eq:kp1}. 

\subsection{Rational solutions of KPI related to the Boussinesq equation}
The Boussinesq equation \eqref{eq:bq} is a symmetry reduction of the KPI equation \eqref{eq:kp1} and so the generalised rational solutions  $\widetilde{u}_n(x,t;\alpha,\beta)$ given by \eqref{untilde} of the Boussinesq equation can be used to generate rational solutions of the KPI equation.
If in the KPI equation \eqref{eq:kp1} we make the travelling wave reduction
\[ v(\xi,\eta,\tau)=u(x,t),\qquad x=\xi-3\tau,\quad t=\eta,\]
then $u(x,t)$ satisfies the Bouss\-inesq equation \eqref{eq:bq}.
Consequently given a solution of the Bouss\-inesq equation \eqref{eq:bq}, then we can derive a solution of the KPI equation \eqref{eq:kp1}. In particular, if 
\[ u(x,t)=2\pderiv[2]{}{x}\ln F(x,t),\]
for some known $F(x,t)$, is a solution of the Bouss\-inesq equation \eqref{eq:bq}, then
\[ v(\xi,\eta,\tau)=2\pderiv[2]{}{\xi}\ln F(\xi-3\tau,\eta),\]
is a solution of the KPI equation \eqref{eq:kp1}. For example the choice $F(x,t)=x^2+t^2+1$ gives the lump solution \eqref{kp:lump} of KPI. 

Using the generalised rational solution $\widetilde{u}_2(x,t;\alpha,\beta)$ \eqref{bq:genrat2} of the Bouss\-inesq equation \eqref{eq:bq} we obtain the rational solution of the KPI equation \eqref{eq:kp1} given by
\begin{equation}\label{kp1sol:bq2} v(\xi,\eta,\tau;\alpha,\beta)=2\pderiv[2]{}{\xi}\ln F_2^{\rm bq}(\xi,\eta,\tau;\alpha,\beta),\end{equation}
where $F_2^{\rm bq}(\xi,\eta,\tau;\alpha,\beta)=\widetilde{F}_2(x,t;\alpha,\beta)$, i.e.
\begin{align}
F_2^{\rm bq}(\xi,\eta,\tau;\alpha,\beta)={\xi}^{6}&-18 \tau{\xi}^{5}+ 3\left( 45 {\tau}^{2}+ {\eta}^{2}+{\tfrac {25}{9}}\right) {\xi}^{4}- 12\left(45 {\tau}^{2}+3 {\eta}^{2}+\tfrac{25}{3}\right)\tau {\xi}^{3}\nonumber\\
&+ \big\{ 3 {\eta}^{4}+ 18\left(9 {\tau}^{2}+\tfrac53 \right) {\eta}^{2}+1215 {\tau}^{4}+450 {\tau}^{2}-{\tfrac {125}{9}} \big\} {\xi}^{2}\nonumber\\
&- \big\{18{\eta}^{4}+ 36\left(9 {\tau}^{2}+5\right) {\eta}^{2}+1458 {\tau}^{4}+900 {\tau}^{2}+{\tfrac {250}{3}} \big\} \tau\xi\nonumber\\
&+{\eta}^{6}+27\left({\tau}^{2}+{\tfrac {17}{81}} \right) {\eta}^{4} + 9\left( 27 {\tau}^{4}+30 {\tau}^{2}+{\tfrac {475}{81}}\right) {\eta}^{2}\nonumber\\
&+729 {\tau}^{6}+675 {\tau}^{4}-125 {\tau}^{2}+{\tfrac {625}{9}} 
+2\a\left\{3  {\xi}^{2}\eta -18 \xi\tau \eta -{\eta}^{3}+ \left( 27 {\tau}^{2}+\tfrac53 \right) \eta\right\}\nonumber\\
&+2\b\left\{{\xi}^{3}-9  {\xi}^{2} \tau- \left(3 {\eta}^{2}-27 {\tau}^{2}+\tfrac13 \right) \xi-27 {\tau}^{3}+9 \tau {\eta}^{2}+\tau\right\}
+ {\a}^{2}+{\b}^{2}.\label{kp1F:bq2}
\end{align}
We remark that this polynomial, in scaled coordinates, is given by Gorshkov, Pelinovsky and Stepanyants \citep{refGoPS}, see their equation (4.2), though the authors don't mention the Boussinesq equation.

\subsection{A more general rational solution}
If we compare the polynomials $F_2^{\rm nls}(\xi,\eta,\tau;\a)$ and $F_2^{\rm bq}(\xi,\eta,\tau;\alpha,\beta)$, respectively given by  \eqref{kp1F:nls2} and \eqref{kp1F:bq2}, then we see that they are fundamentally different. As we shall now demonstrate, they are special cases of a more general polynomial.
Consider the polynomial $\mathcal{F}_2(\xi,\eta,\tau;\mu,\a,\b)$, with parameters $\mu$, $\alpha$ and $\beta$, given by
\begin{align} 
\mathcal{F}_2(\xi,\eta,\tau;\mu,\a,\b)
={\xi}^{6}&-18 \tau {\xi}^{5}+ (3 {\eta}^{2}+ 135 \tau^{2}-6\mu^{2}+9 ) {\xi}^{4} 
- \left\{36{\eta}^{2} +540 \tau^{2} - 12(6\mu^{2}+6\mu-7 )\right\} \tau{\xi}^{3}\nonumber\\
&+ \left\{ 3 {\eta}^{4}+ 18(9\tau^{2}-2\mu+1 ){\eta}^{2}+ 1215 \tau^{4}  -\ 54(6\mu^{2}+12\mu-5) \tau^{2}  
\right.\nonumber\\ &\qquad\left. +\ 9\mu ( \mu+2 )  ( \mu^{2}-2\mu+2 )\right\} {\xi}^{2}\nonumber\\
 &- \left\{18{\eta}^{4} +36 ( 9 \tau^{2}+5){\eta}^{2}+1458 \tau^{4} 
 -324 ( 2\mu^{2}+6\mu -1)\tau^{2}\right.\nonumber\\ &\qquad\left. 
+\ 18\mu ( 3\mu^{3}+12\mu^{2}-2\mu+12 ) 
\right\} \tau\xi 
+ {\eta}^{6}+ (27 \tau^{2}+6\mu^{2}+12\mu+9 ){\eta}^{4}  \nonumber\\
& + \left\{ 243 \tau^{4} + 54(6\mu+7) \tau^{2}+9(\mu^{4}+4\mu^{3}+6\mu^{2}-4\mu+4)\right\}  {\eta}^{2}\nonumber\\ 
&+729 \tau^{6} - 81(\mu^{2}+24\mu-1) \tau^{4} 
+ 9(9{\mu}^{4}+72{\mu}^{3}+150{\mu}^{2}+132\mu+16) \tau^{2}\nonumber\\&
+9( \mu^{2}-2\mu+2 )^{2} 
+2\a\left\{3 \eta {\xi}^{2}-18 \tau \eta \xi-{\eta}^{3}+ 3\left[ 9{\tau}^{2}-\mu( \mu+2)  \right] \eta\right\}\nonumber\\
&+2\b\left\{{\xi}^{3}-9\tau {\xi}^{2}-6({\eta}^{2}-9{\tau}^{2}+{\mu}^{2}) \xi+9\tau {\eta}^{2}-27{\tau}^{3} 
+\ 3( 3{\mu}^{2}+12\mu+4) \tau\right\} \nonumber\\ &
+\a^2 +\beta^2.\label{kp1:FF2}
\end{align}
This polynomial has both the polynomials $F_2^{\rm nls}(\xi,\eta,\tau;\a)$ and $F_2^{\rm bq}(\xi,\eta,\tau;\alpha,\beta)$ as special cases, specifically
\begin{align*}
F_2^{\rm nls}(\xi,\eta,\tau;\a)&=\mathcal{F}_2(\xi,\eta,\tau;1,\a,0), \qquad 
F_2^{\rm bq}(\xi,\eta,\tau;\alpha,\beta)=\mathcal{F}_2(\xi,\eta,\tau;-\tfrac13,\a,\b). 
\end{align*}
Furthermore
\begin{equation}\label{kp1sol:gen2}
v(\xi,\eta,\tau;\mu,\a,\b) = 2\pderiv[2]{}{\xi}\ln\mathcal{F}_2(\xi,\eta,\tau;\mu,\a,\b),
\end{equation} with $\mathcal{F}_2(\xi,\eta,\tau;\mu,\a,\b)$ given by \eqref{kp1:FF2},
is a solution of the KPI equation \eqref{eq:kp1}, which includes as special cases the solutions \eqref{kp1sol:nls2}, when $\mu=1$ and $\b=0$, and \eqref{kp1sol:bq2}, when $\mu=-\tfrac{1}{3}$, as is easily shown.

\begin{figure}[ht]
\[\begin{array}{c@{\quad}c@{\quad}c} 
\includegraphics[width=2in]{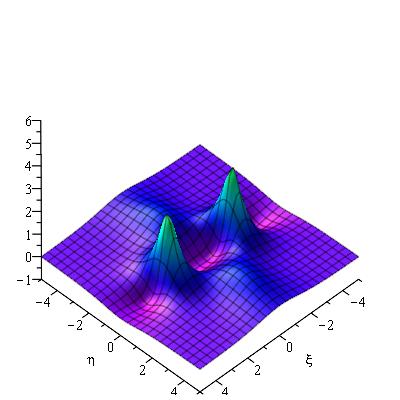} & \includegraphics[width=2in]{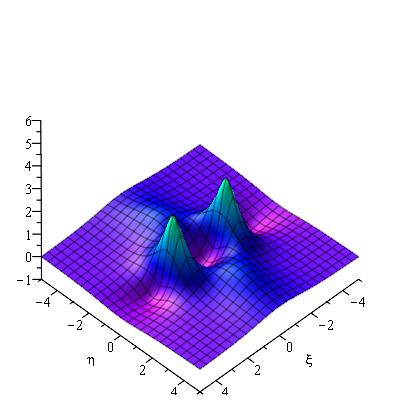} & \includegraphics[width=2in]{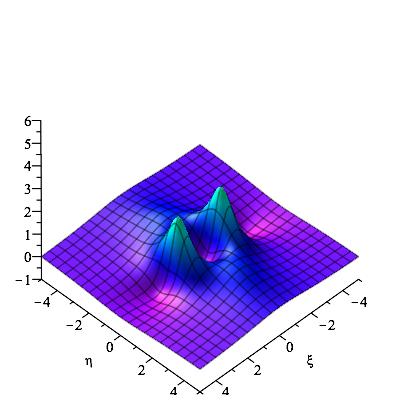}  \\[2pt]
\mu=-1 & \mu=-\tfrac23 & \mu=-\tfrac13\\
\includegraphics[width=2in]{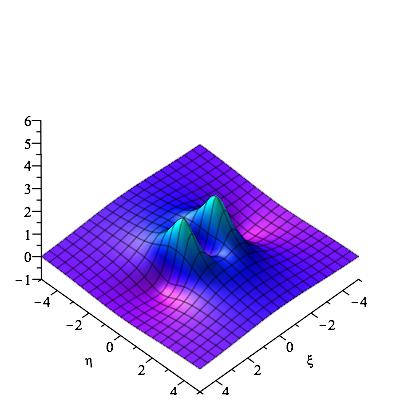} &
 \includegraphics[width=2in]{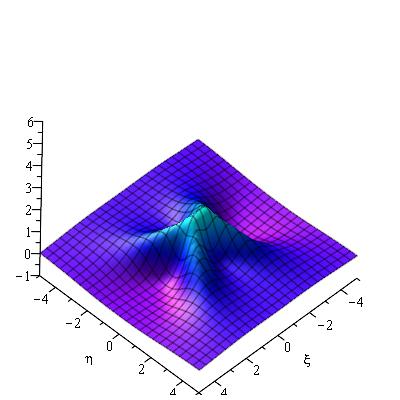}  & \includegraphics[width=2in]{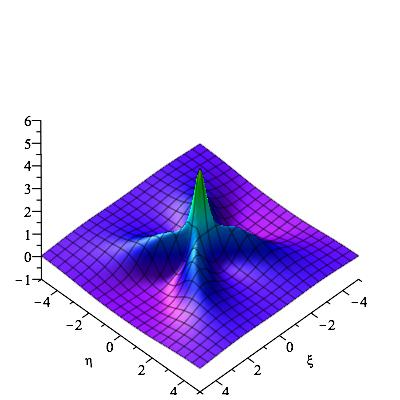} \\
\mu=0 &\mu=0.5115960325&  \mu=0.75 \\
\includegraphics[width=2in]{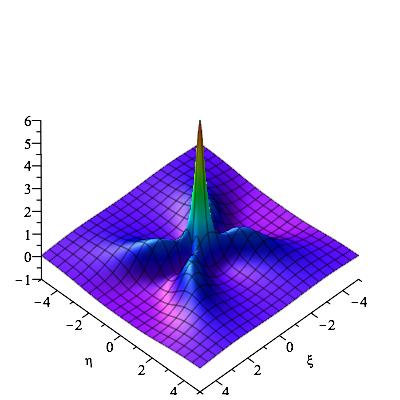} & \includegraphics[width=2in]{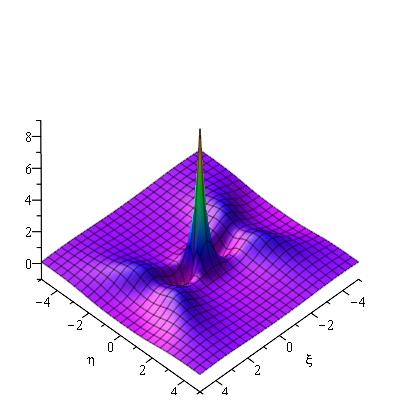}  & \includegraphics[width=2in]{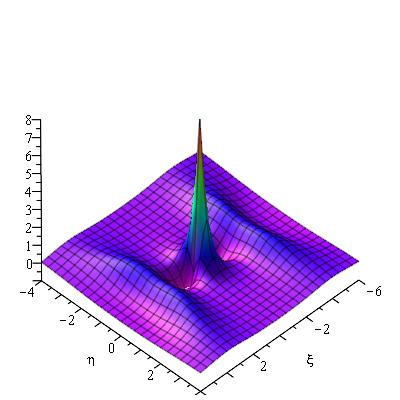} \\
\mu=1 &\mu=\tfrac12(1+\sqrt{5}) & \mu=2\\
\includegraphics[width=2in]{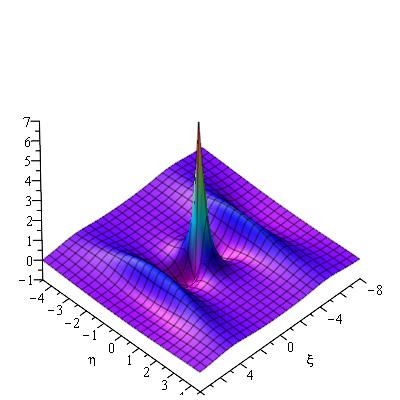} & \includegraphics[width=2in]{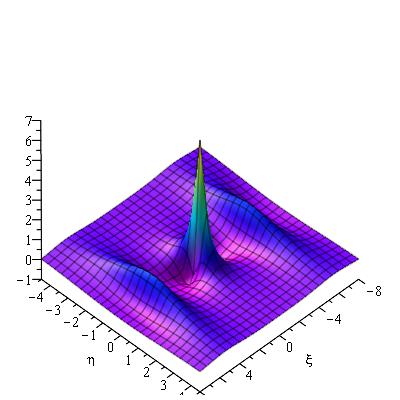}  & \includegraphics[width=2in]{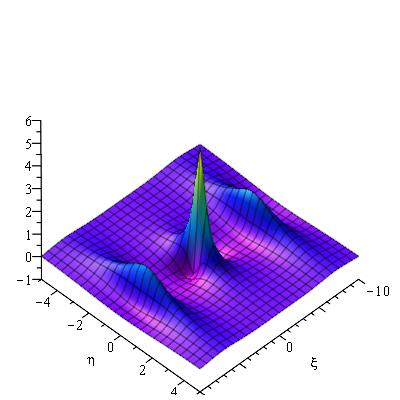} \\
\mu=2.5 &\mu=3 & \mu=4
\end{array}\]
\caption{\label{fig:vgen}The initial solution $v(\xi,\eta,0;\mu,0,0)$ given by \eqref{kp1sol:gen2} is plotted for various choices of the parameter $\mu$. When $\mu=-\tfrac13$ the initial solution corresponds to that arising from the Boussinesq equation \eqref{eq:bq} and when $\mu=1$ to the initial solution from the focusing NLS equation \eqref{eq:fnls}.}
\end{figure}

In Figure \ref{fig:vgen}, the initial solution $v(\xi,\eta,0;\mu,0,0)$ given by \eqref{kp1sol:gen2} is plotted for various choices of the parameter $\mu$. When $\mu=1$, then this arises from the solution \eqref{kp1sol:nls2} derived from the focusing NLS equation \eqref{eq:fnls} whilst when $\mu=-\tfrac13$, then this arises from the solution \eqref{kp1sol:bq2} derived from the Boussinesq equation \eqref{eq:bq}. From Figure \ref{fig:vgen} we can see that for $\mu<\mu^*$, the solution $v(\xi,\eta,0;\mu,0,0)$ has two peaks on the line $\eta=0$, which coalesce when $\mu=\mu^*$ to form one peak at $\xi=\eta=0$. By considering when 
\[\left. \pderiv[2]{}{\xi} v(\xi,0,0;\mu,0,0) \right|_{\xi=0}=-\frac{8(3{\mu}^{4}+12{\mu}^{3}+16{\mu}^{2}-6)}{(\mu^2-2\mu+2)^2}=0,\]
then $\mu^*$ is the real positive root of \[3{\mu}^{4}+12{\mu}^{3}+16{\mu}^{2}-6
=3\left[\mu^2+2(1-\tfrac13\sqrt{6})\mu+2-\sqrt{6}\right]\left[\mu^2+2(1+\tfrac13\sqrt{6})\mu+2+\sqrt{6}\right]=0,\]
i.e.\ $\mu^*=-1+\tfrac13 \sqrt {6}+\tfrac13 \sqrt {-3+3\sqrt {6}}\approx 0.5115960325$. 
For $\mu>\mu^*$, it can be shown that 
\[v(0,0,0;\mu,0,0)=\frac{4\mu(\mu+2)}{\mu^2-2\mu+2},\] 
increases until it reaches a maximum height of $4(2+\sqrt{5})$ when $\mu=\tfrac12(1+\sqrt{5})$, which is the golden mean!
\section{\label{sec5}Discussion}
In this paper we have derived a sequence of algebraically decaying rational solutions of the Boussinesq equation \eqref{eq:bq} which depend on two arbitrary parameters, have an interesting structure and have a similar appearance to rogue-wave solutions in the sense that they have isolated ``lumps". The associated special polynomial, which has equal weight in $x$ and $t$, satisfies a bilinear equation of Hirota type and comprises of a linear combination of four independent solutions of the bilinear equation, something remarkable for a solutions of a bilinear equation. The derivation of a representation of these special polynomials as determinants is currently under investigation and we do not pursue this further here. We remark that other types of exact solutions of the Boussinesq equation \eqref{eq:bq} can be derived using the bilinear equation \eqref{eq:bilin} including breather solutions \citep{refTajMur89,refTajMur91} and rational-soliton solutions \citep{refRLQH}.

Using our rational solutions of the Boussinesq equation \eqref{eq:bq}, we derived rational solutions of the the KPI equation \eqref{eq:kp1} and compared them to those obtained from rational solutions of the focusing NLS equation \eqref{eq:fnls} by Dubard and Matveev \citep{refDubMat11,refDubMat13}. It was shown that the two sets of solutions are fundamentally different and both are special cases of a more general rational solution.
We remark that Ablowitz \etal\ \citep{refACTV,refAV,refVA99} derived a hierarchy of algebraically decaying rational solutions of the KPI equation \eqref{eq:kp1} which have the form
\begin{equation}\label{KPi:ACTV}
v_m(\xi,\eta,\tau)=2\pderiv[2]{}{\xi} \ln G_m(\xi,\eta,\tau),\end{equation}
where $G_m(\xi,\eta,\tau)$ is a polynomial of degree $2m$ in $\xi$, $\eta$ and $\tau$. 
These rational solutions are derived in terms of the eigenfunctions of the non-stationary \sch\ equation
\begin{equation}\label{KPist1}
\i\varphi_{\eta}+\varphi_{\xi\xi}+v\varphi=0,
\end{equation}
with potential $v=v(\xi,\eta,\tau)$, which is used in the solution of KPI \eqref{eq:kp1} by inverse scattering; equation \eqref{eq:kp1} is obtained from the compatibility of \eqref{KPist1} and 
\begin{equation}\label{KPist2}
\varphi_{\tau}+4\varphi_{\xi\xi\xi}+6v\varphi_{\xi}+w\varphi=0,\qquad w_{\xi}=v.
\end{equation}
This is a fundamentally different hierarchy of solutions of the KPI equation \eqref{eq:kp1} compared to those discussed in \S\ref{sec4}, not least because it involves polynomials of all even degrees, not just of degree $n(n+1)$, with $n\in\mathbb{N}$.

\subsection*{Acknowledgment}
PAC thanks Nail Akhmediev, Adrian Ankiewicz and Andrew Bassom for helpful comments and illuminating discussions and the School of Mathematics \& Statistics at the University of Western Australia, Perth, Australia, for their hospitality during his visits when some of this research was done.  We also thank the reviewers for their helpful comments.

\subsection*{Appendix}
\[\begin{split}F_4(x,t)&={x}^{20}+ \left(10{t}^{2} + 90\right) {x}^{18}+ \left(45{t}^{4} +1010{t}^{2}+1845 \right) {x}^{16}
\nonumber\\ &\phantom{={x}^{20}\ }+\left(120{t}^{6}+4600{t}^{4}+30600{t}^{2}+13000 \right) {x}^{14} \nonumber\\ 
&\phantom{={x}^{20}\ }+\left(210{t}^{8}+11480{t}^{6}+151900{t}^{4}+393400{t}^{2}-{\tfrac {2097550}{9}} \right) {x}^{12}
 \nonumber\\ &\phantom{={x}^{20}\ }+\left(252{t}^{10}+17500{t}^{8}+367640{t}^{6}+2095800{t}^{4}+
{\tfrac {11948300}{9}}{t}^{2}+{\tfrac {232696100}{27}} \right) {x}^{10}
 \nonumber\\ &\phantom{={x}^{20}\ }+\left(210{t}^{12}+16940{t}^{10}+501550{t}^{8}+5010600{t}^{6}
+{\tfrac {39702250}{3}}{t}^{4}+{\tfrac {180407500}{9}}{t}^{2} 
-{\tfrac {6596112250}{27}} \right) {x}^{8}
 \nonumber\\ &\phantom{={x}^{20}\ }+\left(120{t}^{14}+10360{t}^{12}+400120{t}^{10}+5601400{t}^{8
}+{\tfrac {141659000}{3}}{t}^{6}+{\tfrac {23569000}{9}}{t}^{4} \right.\nonumber\\&\qquad\qquad\left. 
-\ {\tfrac {19319573000}{27}}{t}^{2}+{\tfrac {86014747000}{27}} \right) {x}^{6}
 \nonumber\\ &\phantom{={x}^{20}\ }+\left(45{t}^{16}+3800{t}^{14}+179900{t}^{12}+3504200{t}^{10}
+{\tfrac {98796250}{3}}{t}^{8}+{\tfrac {1675457000}{9}}{t}^{6}\right.\nonumber\\&\qquad\qquad\left.
-\ {\tfrac {15031607500}{27}}{t}^{4}+{\tfrac {410944625000}{27}}{t}^{2}+
{\tfrac {2352823598125}{81}} \right) {x}^{4}
 \nonumber\\ &\phantom{={x}^{20}\ }+\left(10{t}^{18}+730{t}^{16}+39400{t}^{14}+1320200{t}^{12}+{
\tfrac {74612300}{9}}{t}^{10}+{\tfrac {1165839500}{9}}{t}^{8}\right.\nonumber\\&\qquad\qquad\left.+{
\tfrac {73409791000}{27}}{t}^{6}+\ {\tfrac {1122199715000}{27}}{t}^{4}
+{\tfrac {10744980496250}{81}}{t}^{2}-{\tfrac {8594611821250}{243}}
 \right) {x}^{2}\nonumber\\ &\phantom{={x}^{20}\ }+{t}^{20}+50{t}^{18}+2565{t}^{16}+122200{t}^{14}+{\tfrac {40078850
}{9}}{t}^{12}+{\tfrac {2423740900}{27}}{t}^{10}+{\tfrac {44477105750
}{27}}{t}^{8} \nonumber\\&\qquad\qquad+{\tfrac {177775871000}{9}}{t}^{6}+{\tfrac {
4304738108125}{81}}{t}^{4}+{\tfrac {42895279813750}{243}}{t}^{2}+{\tfrac {73054200480625}{729}}\end{split}\]
\[\begin{split}
F_5(x,t)&=x^{30}+ \left( 15{t}^{2}+{\tfrac {605}{3}} \right) {x}^{28}+ \left( 105{t}^{4}+3290{t}^{2}+12705 \right) {x}^{26}
+ \left( 455{t}^{6}+{\tfrac {71575}{3}}{t}^{4}+{\tfrac {2265725}{9}}{t}^{2}+{\tfrac {25939375}{81}} \right) {x}^{24}
\\ &\qquad\quad+ \left( 1365{t}^{8}+{\tfrac {309260}{3}}{t}^{6}+{\tfrac {17897950}{9
}}{t}^{4}+{\tfrac {26849900}{3}}{t}^{2}+{\tfrac {374564575}{243}}
 \right) {x}^{22}\\ 
&\qquad\quad+ \left( 3003{t}^{10}+298375{t}^{8}+{\tfrac {79208990}{9}}{t}^{6}+
{\tfrac {725413150}{9}}{t}^{4}+{\tfrac {1327947775}{9}}{t}^{2}+{\tfrac {45146222275}{729}} \right) {x}^{20}
\\ &\qquad\quad+ \left( 5005{t}^{12}+{\tfrac {1842610}{3}}{t}^{10}+{\tfrac {74936225
}{3}}{t}^{8}+{\tfrac {30256387700}{81}}{t}^{6}+{\tfrac {416681967625
}{243}}{t}^{4}+{\tfrac {1062878489750}{729}}{t}^{2}\right.\\ &\qquad\qquad\qquad\left. -{\tfrac {29949453408875}{6561}} \right) {x}^{18}
\\ &\qquad\quad+ \left( 6435{t}^{14}+929005{t}^{12}+{\tfrac {145887805}{3}}{t}^{
10}+{\tfrac {9444440425}{9}}{t}^{8}+{\tfrac {716701225625}{81}}{t}^{
6}+{\tfrac {4765327769125}{243}}{t}^{4}\right.\\ &\qquad\qquad\qquad\left.-{\tfrac {16069741485875}{729}}
{t}^{2}+{\tfrac {1572487588700875}{6561}} \right) {x}^{16}
\\ &\qquad\quad+ \left( 6435{t}^{16}+1049400{t}^{14}+{\tfrac {201729500}{3}}{t}^{
12}+{\tfrac {17384033800}{9}}{t}^{10}+{\tfrac {679848919750}{27}}{t}
^{8}+{\tfrac {29329239247000}{243}}{t}^{6}\right.\\ &\qquad\qquad\qquad\left.+{\tfrac {56763015732500}{
729}}{t}^{4}+{\tfrac {877079786275000}{729}}{t}^{2}-{\tfrac {
145319532381244375}{19683}} \right) {x}^{14}
\\ &\qquad\quad+ \left( 5005{t}^{18}+888965{t}^{16}+{\tfrac {201107900}{3}}{t}^{
14}+{\tfrac {7268596300}{3}}{t}^{12}+{\tfrac {397343633750}{9}}{t}^{
10}+{\tfrac {3094794221750}{9}}{t}^{8}\right.\\ &\qquad\qquad\qquad\left.+{\tfrac {877248309206500}{729}}
{t}^{6}+{\tfrac {7522818112617500}{2187}}{t}^{4}-{\tfrac {
338877246089256875}{6561}}{t}^{2}-{\tfrac {1153508042510140625}{
177147}} \right) {x}^{12}
\\ &\qquad\quad+ \left( 3003{t}^{20}+{\tfrac {1682450}{3}}{t}^{18}+{\tfrac {
144448885}{3}}{t}^{16}+{\tfrac {18942077000}{9}}{t}^{14}+
49769993350{t}^{12}+{\tfrac {16141595185100}{27}}{t}^{10}\right.\\ &\qquad\qquad\qquad\left.+{\tfrac {
2217737551163750}{729}}{t}^{8}+{\tfrac {9963380300797000}{729}}{t}^
{6}-{\tfrac {1297656625261390625}{6561}}{t}^{4}\right.\\ &\qquad\qquad\qquad\left.+{\tfrac {
4533029626565151250}{19683}}{t}^{2}+{\tfrac {4174111038326870361875}{
531441}} \right) {x}^{10}
\\ &\qquad\quad+ \left( 1365{t}^{22}+258335{t}^{20}+{\tfrac {73529225}{3}}{t}^{18
}+{\tfrac {11361306425}{9}}{t}^{16}+{\tfrac {976840075750}{27}}{t}^{
14}+{\tfrac {17752164295250}{27}}{t}^{12}\right.\\ &\qquad\qquad\qquad\left.+{\tfrac {3658725849605750}{
729}}{t}^{10}+{\tfrac {3515840993183750}{243}}{t}^{8}-{\tfrac {
195785332934489375}{729}}{t}^{6}\right.\\ &\qquad\qquad\qquad\left.+{\tfrac {24978207925819946875}{6561}
}{t}^{4}+{\tfrac {1159166663661630903125}{19683}}{t}^{2}+{\tfrac {
4904143764303914178125}{531441}} \right) {x}^{8}
\\ &\qquad\quad+ \left( 455{t}^{24}+{\tfrac {251020}{3}}{t}^{22}+{\tfrac {76657070}{
9}}{t}^{20}+{\tfrac {41289423700}{81}}{t}^{18}+{\tfrac {
1362831787625}{81}}{t}^{16}+{\tfrac {98913216479000}{243}}{t}^{14}\right.\\ &\qquad\qquad\qquad\left.+
{\tfrac {4366923310634500}{729}}{t}^{12}+{\tfrac {14325694558021000}{
729}}{t}^{10}+{\tfrac {164980602695610625}{729}}{t}^{8} 
+{\tfrac {38543006652688037500}{2187}}{t}^{6}\right.\\ &\qquad\qquad\qquad\left.
+{\tfrac {12142620899858806568750}
{59049}}{t}^{4}+{\tfrac {84368406785489229287500}{177147}}{t}^{2}
+{\tfrac {1033632925475218502809375}{4782969}} \right) {x}^{6}
\\ &\qquad\quad+ \left( 105{t}^{26}+{\tfrac {53375}{3}}{t}^{24}+{\tfrac {16915150}{9
}}{t}^{22}+{\tfrac {1171587550}{9}}{t}^{20}+{\tfrac {1229272389625}{
243}}{t}^{18}+{\tfrac {32275315890125}{243}}{t}^{16}\right.\\ &\qquad\qquad\qquad\left.+{\tfrac {
2128271542512500}{729}}{t}^{14}+{\tfrac {110365606933697500}{2187}}
{t}^{12}+{\tfrac {6125181130562869375}{6561}}{t}^{10}\right.\\ &\qquad\qquad\qquad\left.+{\tfrac {
184494438219511371875}{6561}}{t}^{8}+{\tfrac {18829554428932184918750
}{59049}}{t}^{6}+{\tfrac {30319073658670395156250}{19683}}{t}^{4}\right.\\ &\qquad\qquad\qquad\left.+{
\tfrac {767901026020862022953125}{531441}}{t}^{2}-{\tfrac {
37763631956445485447328125}{14348907}} \right) {x}^{4}\\ 
&\qquad\quad+ \left( 15{t}^{28}+2170{t}^{26}+{\tfrac {2043125}{9}}{t}^{24}+{
\tfrac {163177700}{9}}{t}^{22}+{\tfrac {8631985775}{9}}{t}^{20}+{
\tfrac {17793313441750}{729}}{t}^{18}\right.\\ &\qquad\qquad\qquad\left.+{\tfrac {584377965527125}{729}}
{t}^{16}+{\tfrac {7043820768985000}{243}}{t}^{14}+{\tfrac {
6235281337588043125}{6561}}{t}^{12}\right.\\ &\qquad\qquad\qquad\left.+{\tfrac {437562641832806971250}{
19683}}{t}^{10}+{\tfrac {5034320101951909278125}{19683}}{t}^{8}+{
\tfrac {296816181647178511587500}{177147}}{t}^{6}\right.\\ &\qquad\qquad\qquad\left.-{\tfrac {
305501861525583991296875}{531441}}{t}^{4}+{\tfrac {
139014074702059270656250}{531441}}{t}^{2} 
+{\tfrac {634083161524687235258734375}{43046721}} \right) {x}^{2}\\&\qquad\quad
+{t}^{30}+{\tfrac {325}{3}}{t}^{28}+10185{t}^{26}+{\tfrac {71587775}{
81}}{t}^{24}+{\tfrac {16294723375}{243}}{t}^{22}+{\tfrac {
2934806885675}{729}}{t}^{20} 
+{\tfrac {1145785364618125}{6561}}{t}^{18} \\ &\qquad\qquad\qquad
+{\tfrac {44166106891704875}{6561}}{t}^{16}+{\tfrac {
4108707388089775625}{19683}}{t}^{14} 
+{\tfrac {774149365283245634375}{177147}}{t}^{12}
\\ &\qquad\qquad\qquad
+{\tfrac {24580063449195140376875}{531441}}{t}^{10}
+{\tfrac {266920437967411700828125}{531441}}{t}^{8} 
+{\tfrac {18940589955229082293759375}{4782969}}{t}^{6} \\ &\qquad\qquad\qquad
+{\tfrac {196432003698991651589796875}{14348907}}{t}^{4} 
+{\tfrac {1654599020642266683930859375}{43046721}}{t}^{2} 
+{\tfrac {293277952222570147203765625}{43046721}}
\end{split}\]

\[\begin{split}
P_3(x,t)&=7{x}^{12}- \left(14{t}^{2}-210 \right) {x}^{10}- \left(63{t}^{4} +630{t}^{2}-{\tfrac {875}{3}}\right) {x}^{8}
-\left(36{t}^{6}+2044{t}^{4}+{\tfrac {16100}{3}}{t}^{2}-{\tfrac {16100}{3}}\right) {x}^{6}\nonumber\\
&\phantom{=7{x}^{12}\ }+ \left( 25{t}^{8}+260{t}^{6}-{\tfrac {39550}{3}}{t}^{4} -{\tfrac {91700}{3}}{t}^{2} -{\tfrac {1066975}{9}} \right) {x}^{4}\nonumber\\
&\phantom{=7{x}^{12}\ }+ \left(18{t}^{10}+{\tfrac {1310}{3}}{t}^{8}+{\tfrac {26140}{3}}{t}^{6}+{\tfrac {146300}{3}}{t}^{4} +{\tfrac {1835050}{9}}{t}^{2}+ {\tfrac {32655350}{27}} \right) {x}^{2}\nonumber\\
&\phantom{=7{x}^{12}\ }-{t}^{12}-\tfrac{10}3{t}^{10}+25{t}^{8}-{\tfrac {1900}{3}}{t}^{6}-{\tfrac {1230775}{9}}{t}^{4} -{\tfrac {2070250}{3}}{t}^{2} +{\tfrac {32680375}{81}}\end{split}\]
\[\begin{split}
Q_3(x,t)&= {x}^{12}- \left(18{t}^{2}-{\tfrac {74}{3}} \right) {x}^{10}-\left( 25{t}^{4} + {\tfrac {1870}{3}}{t}^{2}+{\tfrac {275}{3}}\right) {x}^{8} 
+ \left( 36{t}^{6}-580{t}^{4}-{\tfrac {8860}{3}}{t}^{2}+{\tfrac {4700}{3}} \right) {x}^{6}\nonumber\\
&\phantom{={x}^{12}\ }+ \left( 63{t}^{8}+1820{t}^{6}-{\tfrac {2450}{3}}{t}^{4}-{\tfrac {37100}{3}}{t}^{2}-{\tfrac {247625}{9}} \right) {x}^{4}\nonumber\\
&\phantom{={x}^{12}\ }+ \left(14{t}^{10}+630{t}^{8}+{\tfrac {49700}{3}}{t}^{6}+48300{t}^{4}  +{\tfrac {1877750}{9}}{t}^{2}+ {\tfrac {2898350}{9}}\right) {x}^{2}\nonumber\\
&\phantom{={x}^{12}\ }-7{t}^{12}-98{t}^{10} -{\tfrac {5075}{3}}{t}^{8} -23100{t}^{6}-{\tfrac {2108225}{9}}{t}^{4}-{\tfrac {43900150}{27}}{t}^{2}-{\tfrac {4998175}{81}}\end{split}\]
\[\begin{split}
P_4(x,t)&=9{x}^{20}- \left(30{t}^{2}-770 \right) {x}^{18}- \left(243{t}^{4}+3390{t}^{2}-14245 \right) {x}^{16}
- \left(360{t}^{6}+24360{t}^{4}+107800{t}^{2}-{\tfrac {754600}{9}} \right) {x}^{14}
\\ &\phantom{=9{x}^{20}\ }+ \left( 130{t}^{8}-23720{t}^{6}-{\tfrac {2278220}{3}}{t}^{4}-{
\tfrac {4419800}{3}}{t}^{2}-{\tfrac {51285850}{27}} \right) {x}^{12}
\\ &\phantom{=9{x}^{20}\ }+ \left( 780{t}^{10}+{\tfrac {94820}{3}}{t}^{8}-{\tfrac {759640}{3}}{t}^{6}-{\tfrac {82510120}{9}}{t}^{4}
+{\tfrac {16762900}{27}}{t}^{2}+{\tfrac {5563180700}{81}} \right) {x}^{10}\\ 
&\phantom{=9{x}^{20}\ }+ \left( 690{t}^{12}+58700{t}^{10}+{\tfrac {3917450}{3}}{t}^{8}+{
\tfrac {79849000}{9}}{t}^{6}-{\tfrac {1064659750}{27}}{t}^{4}-{\tfrac {5795597500}{27}}{t}^{2}-{\tfrac {1367658734750}{729}} \right) {x}^{8}\\ 
&\phantom{=9{x}^{20}\ }+ \left( 152{t}^{14}+{\tfrac {65800}{3}}{t}^{12}+{\tfrac {11986520}{9}}{t}^{10}+{\tfrac {1202215000}{81}}{t}^{8}+{\tfrac {8625185800}{81}
}{t}^{6}+{\tfrac {69758781400}{243}}{t}^{4}\right.\\ &\qquad\qquad\qquad\left. +{\tfrac {1077743975000}{243}}{t}^{2}+{\tfrac {55941010279000}{2187}} \right) {x}^{6}\\ 
&\phantom{=9{x}^{20}\ }- \left(75{t}^{16}+{\tfrac {10360}{3}}{t}^{14}+{\tfrac {66500}{9}}{t}^{12}-{\tfrac {69057800}{9}}{t}^{10}-{\tfrac {20996610250}{243}}
{t}^{8}-{\tfrac {275830555000}{243}}{t}^{6}\right.\\ &\qquad\qquad\qquad\left.
-{\tfrac {5620866905500}{729}}{t}^{4}+{\tfrac {9843829765000}{729}}{t}^{2}-{\tfrac {404610075244375}{2187}} \right) {x}^{4}\\ 
&\phantom{=9{x}^{20}\ }- \left(30{t}^{18}+1790{t}^{16}+{\tfrac {803320}{9}}{t}^{14}+{
\tfrac {28869400}{9}}{t}^{12}+{\tfrac {1473856300}{27}}{t}^{10}+{\tfrac {629478426500}{729}}{t}^{8}
\right.\\ &\qquad\qquad\qquad\left. +{\tfrac {11032069279000}{729}}{t}^{6}+{\tfrac {133702667483000}{729}}{t}^{4}+{\tfrac {112127684226250}{
243}}{t}^{2}+{\tfrac {5297582110686250}{19683}} \right) {x}^{2}\\
&\phantom{=9{x}^{20}\ }+{t}^{20}+{\tfrac {70}{3}}{t}^{18}+{\tfrac {1855}{3}}{t}^{16}+{\tfrac 
{1899800}{81}}{t}^{14}+{\tfrac {438095350}{243}}{t}^{12}+{\tfrac {
88186059500}{729}}{t}^{10}+{\tfrac {14094153477250}{6561}}{t}^{8}\\ &\qquad\qquad\qquad +{
\tfrac {138847640239000}{6561}}{t}^{6}-{\tfrac {823906531765625}{19683
}}{t}^{4}-{\tfrac {20487539830546250}{177147}}{t}^{2}+{\tfrac {266883842659905625}{531441}}
\end{split}\]\[\begin{split}
Q_4(x,t)&={x}^{20}- \left(30{t}^{2}-{\tfrac {230}{3}} \right) {x}^{18}- \left(75{t}^{4}+2830{t}^{2}-{\tfrac {2695}{3}} \right) {x}^{16}
+ \left( 152{t}^{6}-{\tfrac {18680}{3}}{t}^{4}-{\tfrac {615640}{9}}{t}^{2}+{\tfrac {237160}{81}} \right) {x}^{14}\\ 
&\phantom{={x}^{20}\ }+ \left( 690{t}^{8}+{\tfrac {63560}{3}}{t}^{6}-{\tfrac {1276100}{9}}
{t}^{4}-{\tfrac {1832600}{3}}{t}^{2}-{\tfrac {60772250}{243}}\right) {x}^{12}\\ 
&\phantom{={x}^{20}\ }+ \left( 780{t}^{10}+62860{t}^{8}+{\tfrac {8213240}{9}}{t}^{6}-
586600{t}^{4}+{\tfrac {116801300}{27}}{t}^{2}+{\tfrac {8356925500}{729}} \right) {x}^{10}\\ 
&\phantom{={x}^{20}\ }+ \left( 130{t}^{12}+{\tfrac {117700}{3}}{t}^{10}+{\tfrac {4962650}{3
}}{t}^{8}+{\tfrac {1352661800}{81}}{t}^{6}+{\tfrac {179352250}{243}}
{t}^{4}-{\tfrac {117373448500}{729}}{t}^{2} 
-{\tfrac {2249680490750}{6561}} \right) {x}^{8}\\ 
&\phantom{={x}^{20}\ }- \left(360{t}^{14}+15400{t}^{12}-{\tfrac {1286600}{3}}{t}^{10}-
{\tfrac {82943000}{9}}{t}^{8}-{\tfrac {8897211400}{81}}{t}^{6}-{
\tfrac {65165639000}{243}}{t}^{4}\right.\\ &\qquad\qquad\qquad\left.-{\tfrac {3029653781000}{729}}{t}^{
2}-{\tfrac {31684368485000}{6561}} \right) {x}^{6}
\\ &\phantom{={x}^{20}\ }- \left(243{t}^{16}+19560{t}^{14}+{\tfrac {2325260}{3}}{t}^{12}+
{\tfrac {29280440}{9}}{t}^{10}+{\tfrac {59473750}{27}}{t}^{8}-{\tfrac {61024825400}{243}}{t}^{6}\right.\\ &\qquad\qquad\qquad\left.+{\tfrac {320631426500}{729}}{t}^{4}
+{\tfrac {37020326189000}{729}}{t}^{2}-{\tfrac {425463980932375}{19683}} \right) {x}^{4}
\\ &\phantom{={x}^{20}\ }- \left(30{t}^{18}+2910{t}^{16}+191800{t}^{14}+{\tfrac {23294600
}{3}}{t}^{12}+{\tfrac {2501523500}{27}}{t}^{10}+{\tfrac {34775471500}{27}}{t}^{8}\right.\\ &\qquad\qquad\qquad\left.+{\tfrac {1233890119000}{81}}{t}^{6}+{\tfrac {
9121143955000}{81}}{t}^{4}+{\tfrac {114976450146250}{243}}{t}^{2}+{\tfrac {15307611409956250}{177147}} \right) {x}^{2}\\
&\phantom{={x}^{20}\ }+ 9{t}^{20}+370{t}^{18}+15325{t}^{16}+{\tfrac {5192600}{9}}{t}^{
14}+{\tfrac {475032950}{27}}{t}^{12}+{\tfrac {35488950700}{81}}{t}^{
10}+{\tfrac {6433079133250}{729}}{t}^{8}\\ &\qquad\qquad\qquad+{\tfrac {234408499325000}{
2187}}{t}^{6}+{\tfrac {315145911994375}{2187}}{t}^{4}+{\tfrac {
4908421805113750}{19683}}{t}^{2}+{\tfrac {140298620844930625}{531441}
}
\end{split}\]

\def\OUP{O.U.P.} 
\def\CUP{C.U.P.}

\def\refjl#1#2#3#4#5#6#7{\vspace{-0.25cm}
\bibitem{#1}{\frenchspacing\rm#2}, {\rm#6}, 
\textit{\frenchspacing#3}, \textbf{#4}\ (#7)\ #5.}

\def\refbk#1#2#3#4#5{\vspace{-0.25cm}
\bibitem{#1}{\frenchspacing\rm#2}, ``\textit{#3}," #4 (#5).} 

\def\refcf#1#2#3#4#5#6{\vspace{-0.25cm}
\bibitem{#1} \textrm{\frenchspacing#2}, \textrm{#3},
in ``\textit{#4}" [{\frenchspacing#5}], #6.}

\end{document}